\documentclass{emulateapj-rtx4}
\usepackage{natbib}
\usepackage{epsf,graphics}
\usepackage{subfigure,graphicx}
\usepackage{bm}
\begin{document}

\title{Inflation that runs naturally: gravitational waves and suppression of power at large and small 
scales}

\author{Quinn E. Minor}
\affiliation{Department of Science, Borough of Manhattan Community College, 
City University of New York, New York, NY 10007, USA}
\affiliation{Department of Astrophysics, American Museum of Natural History, 
New York, NY 10024, USA}
\author{Manoj Kaplinghat}
\affiliation{Department of Physics and Astronomy, University of California, 
Irvine CA 92697, USA}

\begin{abstract}
We point out three correlated predictions of the axion monodromy inflation 
model: large amplitude of gravitational waves, suppression of power on horizon 
scales and on scales relevant for the formation of dwarf galaxies. While these 
predictions are likely generic to models with oscillations in the inflaton 
potential, the axion monodromy model naturally accommodates the required 
running spectral index through Planck-scale corrections to the inflaton 
potential.  Applying this model to a combined data set of Planck, ACT, SPT, and 
WMAP low-$\ell$ polarization cosmic microwave background (CMB) data, we find a 
best-fit tensor-to-scalar ratio $r_{0.05} = 0.07^{+0.05}_{-0.04}$ due to 
gravitational waves, which may have been observed by the BICEP2 experiment.  
Despite the contribution of gravitational waves, the total power on large 
scales (CMB power spectrum at low multipoles) is lower than the standard 
$\Lambda$CDM cosmology with a power-law spectrum of initial perturbations and 
no gravitational waves, thus mitigating some of the tension on large scales.  
There is also a reduction in the matter power spectrum of 20-30\% at scales 
corresponding to $k = 10~{\rm Mpc}^{-1}$, which are relevant for dwarf galaxy 
formation. This will alleviate some of the unsolved small-scale structure 
problems in the standard $\Lambda$CDM cosmology. The inferred matter power 
spectrum is also found to be consistent with recent Lyman-$\alpha$ forest data, 
which is in tension with the Planck-favored $\Lambda$CDM model with power-law 
primordial power spectrum.
\end{abstract}

\keywords{cosmology: cosmic background radiation, cosmology: inflation, 
galaxies: dwarf\vspace{1.0mm}}

\section{Introduction}\label{sec:intro}

The Planck mission has established that the cosmic microwave background 
anisotropies on small angular scales are well described by the standard 
$\Lambda$CDM cosmology with a nearly scale-invariant power-law spectrum 
(\citealt{planck2013}). At large scales ($\ell < 40$), however, there appears 
an overall deficit of power compared to what is expected in the benchmark 
$\Lambda$CDM model (\citealt{planck2013c}). The unusual shape and amplitude of 
the power spectrum at low multipoles was first observed by the WMAP mission 
(\citealt{spergel2003}; \citealt{hinshaw2003}) and remains unexplained, 
although hypotheses include a running spectral index (\citealt{feng2003}; 
\citealt{bastero2003}; \citealt{chung2003}; \citealt{kawasaki2003}; 
\citealt{hunt2007}), a breakdown of slow-roll inflation or pre-slow roll phase 
(\citealt{peiris2003}; \citealt{contaldi2003}; \citealt{mortonson2009}; 
\citealt{hazra2014}; \citealt{lello2014}), a contracting pre-inflation phase 
(\citealt{piao2004}), open inflation (\citealt{white2014}), a non-Bunch-Davies 
initial vacuum state (\citealt{ashoorioon2014}), or the presence of an extra 
neutrino species (\citealt{dvorkin2014}; \citealt{anchordoqui2014}).

The tension at low $\ell$ is exacerbated significantly if there exists a 
stochastic gravitational wave background with tensor-to-scalar ratio $r \gtrsim 
0.1$, since tensor perturbations add to the expected CMB temperature 
anisotropies at low multipoles (\citealt{smith2014}). Such a large $r$ has been 
suggested by the BICEP2 experiment in its detection of B-mode polarization in 
the sky at degree-angular scales (\citealt{bicep2014}).  Under the assumption 
that the observed B-mode anisotropies are sourced by primordial gravitational 
waves, they infer $r_{0.05}=0.2_{-0.05}^{+0.07}$. By comparison, from the $TT$ 
spectrum alone, \cite{planck2013} inferred $r_{0.05} < 0.135$ at the 95\% 
confidence level, in tension with the BICEP2 result. At present, the BICEP2 
result is highly uncertain due to the likely presence of contamination by 
foreground dust polarization in the observed field-of-view 
(\citealt{flauger2014}; \citealt{mortonson2014}; \citealt{planck2014}).  
Nevertheless, large-field inflation models---including the simplest chaotic 
models---predict at minimum $r \gtrsim 0.01$ (\citealt{lyth1997}) and thus 
would increase the apparent tension with $\Lambda$CDM.

The simplest way to accommodate the deficit of power at low $\ell$ is to allow 
for a running spectral index, thus departing from a power-law spectrum. When a 
constant running of the spectral index $\alpha =dn_s/d\ln k$ is allowed, 
\cite{planck2013b} finds $\alpha = -0.011\pm 0.008$ in the case $r=0$, where 
the preference for negative running is driven largely by the temperature 
likelihood at low multipoles. More significantly, running also allows for a 
higher tensor contribution, leading to $r_{0.05} < 0.275$ when both running and 
tensors are allowed.

Such a large constant running is difficult to implement in the underlying 
inflation model, since it yields an insufficient number of e-foldings to solve 
the horizon problem (\citealt{easther2006}). Within the context of single-field 
slow-roll inflation, the only way to achieve the 50-60 remaining e-foldings 
necessary after the mode $k=0.05$ Mpc$^{-1}$ leaves the horizon, is if the 
running diminishes or turns positive at larger $k$. Plausible mechanisms exist 
for the running to diminish to zero at larger $k$, for example through 
radiative corrections (\citealt{ballesteros2008}; \citealt{ballesteros2014}) or 
GUT symmetry breaking (\citealt{hazra2014a}); this would, however, imply a 
special scale at which the running ``turns over'' and becomes small, a scale 
comparable to or just smaller than scales observable in the CMB, which would 
seem a remarkable coincidence.  Another, perhaps more natural possibility is 
that the power spectrum oscillates, implying that the inflaton potential may 
contain an oscillatory component. Since it would seem 
unnatural for only one such oscillation to occur during the course of 
inflation, the intriguing possibility arises that the inflaton potential 
contains a gentle oscillation which may occur all the way to the end of 
inflation.  In fact, many large-field models that include oscillations in a 
natural way have been investigated (e.g. \citealt{ashoorioon2006};
(\citealt{silverstein2008}; \citealt{kaloper2009}; \citealt{mcallister2010}; 
\citealt{kaloper2011}; \citealt{kobayashi2011}).

In this article we will test axion monodromy models against the CMB temperature 
anisotropy spectrum, but our results are broadly applicable to inflationary 
potentials with gentle oscillations. We will assume that the oscillation scale 
(in $\log k$) is ``long'', i.e.~comparable to the range of multipoles observed 
in the CMB, naturally leading to a running spectral index as described above.  
We will show that the best-fit model gives three correlated predictions: 1) a 
significant gravitational wave amplitude of order $r \sim 0.1$; 2) a reduction 
of power at large scales (low $\ell$) despite the tensor contribution, thus 
mitigating the existing tension at large scales; 3) there is a corresponding 
significant suppression of power at small scales, particularly at the scales 
relevant to dwarf galaxy formation, which will alleviate some of the 
small-scale structure problems of $\Lambda$CDM.  Finally, although axion 
monodromy allows for a gravitational wave background, we will find that the 
e-folding requirement surprisingly places an upper bound on the 
tensor-to-scalar ratio ($r \lesssim 0.2$), which is in tension with the BICEP2 
measurement unless a significant portion of the observed B-mode signal is due 
to foreground contamination rather than primordial gravitational waves.

The paper is outlined as follows. In Section \ref{sec:background} we discuss 
the theoretical motivation for axion monodromy models and derive formulae for 
the power spectra for scalar and tensor perturbations. In Section 
\ref{sec:priors} we discuss our choice of parameters and the priors for each. 
The resulting constraints are shown in Section \ref{sec:results}; the 
constraints on the oscillation parameters are discussed in Section 
\ref{sec:axion_constraints}, the best-fit model is discussed in Section 
\ref{sec:bestfit}, and the constraint on the tensor-to-scalar ratio $r$ is 
discussed in Section \ref{sec:nr_constraints}.  In Sections \ref{sec:lowl} and 
\ref{sec:highl} we compare our model to the usual constant-running model at low 
and high multipoles, respectively. In Section \ref{sec:amending_efoldings} we 
discuss whether our model can be amended to allow for a higher tensor-to-scalar 
ratio.  Section \ref{sec:smallscale_probs} investigates the implications of an 
oscillating power spectrum for the small-scale problems of $\Lambda$CDM: the 
effect on dwarf galaxy formation is discussed in Section \ref{sec:dwarfs}, 
while in Section \ref{sec:lyman_alpha} our results are compared to recent 
Lyman-$\alpha$ forest data and the prospects for other small-scale probes of 
the matter power spectrum are discussed.  We conclude with a summary of our 
main points in Section \ref{sec:conclusions}.

\section{Theoretical Background}\label{sec:background}

\subsection{Motivation for large-field inflation models}

During slow roll inflation, it is easily shown that a relation exists between 
the tensor-to-scalar ratio $r$ and the overall shift $\Delta\phi$ in the scalar 
field from CMB scales to the end of inflation:

\begin{equation}
\frac{\Delta\phi}{M_p} = \mathcal{O}(1)\times\left(\frac{r}{0.01}\right)^{1/2}
\label{eq:lyth_bound}
\end{equation}
where $M_p=\left(8\pi G\right)^{-1/2}$ is the reduced Planck mass. The 
importance of this relation, known as the Lyth bound (\citealt{lyth1997}), is 
that a significant gravitational wave contribution $r \gtrsim 0.01$ implies 
that the scalar field value changes by more than the Planck energy $M_p$. These 
so-called large-field models imply that any fields coupled to the inflaton with 
at least gravitational strength will receive corrections to the coupling 
strengths, resulting in an infinite series of Planck-suppressed terms 
contributing to the scalar field potential.  Such terms spoil the flatness of 
the potential, inhibiting inflation, unless there is an exquisite degree of 
fine-tuning in the coefficients. The only way to avoid fine-tuning is if a 
symmetry ``protects'' the potential from large contributions---in particular, 
an approximate shift symmetry $\phi \rightarrow \phi + a$ in the corresponding 
Lagrangian ensures that the correction to the potential is \emph{at worst} 
periodic, and thus its magnitude can in principle be controlled.

While many different potentials can approximately satisfy the above mentioned 
shift symmetry, there is no guarantee that such a potential admits a UV 
completion in quantum gravity; indeed, it is far more likely that the class of 
potentials which can be derived from a corresponding UV-complete theory obeying 
the same approximate shift symmetry is relatively restricted. For this reason, 
merely considering generic renormalizable effective field theories is not 
sufficient for large-field models. Instead, one should consider inflation 
models that can be derived from quantum gravity, with string theory being the 
best developed to date.  Although a few alternatives exist 
(\citealt{freese1990}; \citealt{dimopoulos2008}; \citealt{wan2014}; 
\citealt{neupane2014a}), one of the best-motivated of these models is axion 
monodromy (\citealt{silverstein2008}; \citealt{mcallister2010}; 
\citealt{flauger2010}; \citealt{aich2013}).  Axion fields $a$ in string theory 
naturally obey a discrete symmetry $a \rightarrow a + 2\pi$.  The phenomenon of 
monodromy appears when axions are coupled to fluxes in a compactified 
higher-dimensional space (in the string theoretic description, this occurs in 
the presence of a wrapped brane in a Calabi-Yau manifold). The shift symmetry 
is slightly broken, allowing the field potential energy to change by a large 
amount as one traverses many cycles in the compactified space, while all the 
remaining microphysics is periodic in field space; this is analogous to a 
spiral staircase where the overall height can change without bound through many 
cycles.

The resulting potential for axion monodromy models consists of a monomial 
term plus a sinusoidal term:

\begin{equation}
V(\phi) = \lambda \phi^p + A\sin\left(\frac{\phi}{f} + \psi\right)
\label{potential_original}
\end{equation}
where we are now working in Planckian units where $M_p=\left(8\pi 
G\right)^{-1/2}=1$ for the remainder of this paper.  The parameter $f$ 
corresponds to the period of oscillation and is known as the \emph{axion decay 
constant}.  Although in its original incarnation the monodromy potential has 
$p=1$, depending on the flux coupling and brane configuration one can also 
achieve other discrete values such as $p=2/3$, $p=2$, $p=3$ and so on 
(\citealt{mcallister2014}).  Since we are interested here in what $p$-value(s) 
are preferred by CMB data, we will vary $p$ as a free parameter, with the 
ansatz that $p$ can be any positive real number.

\subsection{Axion monodromy power spectrum}\label{sec:power_spectrum}

For fitting purposes, we find it useful to make the transformation $\phi 
\rightarrow \phi_{min} - \phi$, where $\phi_{min}$ will be chosen so that $\phi 
= 0$ corresponds to the CMB pivot scale $k_* = 0.05$ Mpc$^{-1}$. With a 
suitable relabeling of parameters, we can recast the potential in the following 
form,

\begin{equation}
\frac{V}{V_*} = \left(1-a\sin\delta\right)\left(1-\phi/\phi_{min}\right)^p + a \sin\left(\frac{\phi}{f} + \delta\right)
\label{potential}
\end{equation}
which is written so that $V(\phi=0) = V_*$. We choose $\phi_{min}$ to be 
positive so the field rolls in the positive direction, i.e.~$\phi$ increases as 
it rolls.

To find the primordial power spectra for scalar and tensor perturbations, one 
typically uses the slow roll approximation, which assumes the Hubble parameter 
varies slowly enough that the quasi-de Sitter solution of the equation for 
quantum perturbations can be used (\citealt{mukhanov1992}; for a review see 
\citealt{baumann2009}).  Careful consideration is required here, however, 
because slow roll can break down if oscillations in the potential are 
sufficiently rapid, in which case the relevant equation must be solved 
directly.  This occurs if the amplitude is sufficiently large, or if the axion 
decay constant $f$ is small (\citealt{flauger2010}). Since in this paper we 
focus only on long-wavelength oscillations in the power spectrum, the slow roll 
approximation holds very well in the parameter region of interest. The proof of 
this is given in appendix \ref{sec:slowroll} for the interested reader.

The power spectrum for curvature perturbations in the slow-roll approximation 
is given by $\Delta_{\mathcal{R}}^2 \sim \frac{H^2}{\epsilon} \sim 
\frac{V^3}{V_{,\phi}^{2}}$.  Normalizing the power spectrum by the scalar 
amplitude at the pivot scale $A_s$, we find

\begin{eqnarray}
\label{powerspec}
&&\Delta_{\mathcal{R}}^2 = A_s \mathcal{N}^2 \times \nonumber \\ 
&&\frac{\left[(1-a\sin\delta)(1-\phi_k/\phi_{min})^p + a \sin(\frac{\phi_k}{f} + \delta)\right]^3}{\left[(1-a\sin\delta)(1-\phi_k/\phi_{min})^{p-1} - \left(\frac{\phi_{min}}{p f}\right) a \cos(\frac{\phi_k}{f} + \delta)\right]^2}, \nonumber \\
\end{eqnarray}
where $\phi_k$ corresponds to the scalar field value at the time when the mode 
with wavenumber $k$ left the horizon (to be determined shortly), and
\begin{equation}
\label{nfactor}
\mathcal{N} = 1 - a\sin\delta - \left(\frac{\phi_{min}}{nf}\right) a\cos\delta.
\end{equation}

Note that at the pivot scale ($\phi_k = 0$), this reduces to 
$\Delta_{\mathcal{R}}^2 = A_s$ as it should.

Meanwhile, the primordial tensor power spectrum is given by $\Delta_t^2 \sim 
H^2 \sim V$, and since it is normalized by $A_t = r A_s$ where $r$ is the 
tensor-to-scalar ratio at the pivot scale $k_*$, we have

\begin{equation}
\label{tensor_powerspec}
\Delta_t^2 = r A_s \frac{V(\phi_k)}{V_*}
\end{equation}
where $V/V_*$ is given in equation \ref{potential}.\footnote{Note that the 
scale of the potential $V_*$ will not enter into our equations directly, since 
we only encounter the combination $V/V_*$.  Nevertheless, $V_*$ is not an 
independent parameter but is rather determined by the normalization of 
$\Delta_t^2$, with the result $V_* = \frac{3\pi^2}{2}rA_s$ in the slow roll 
approximation.}

Finally, we still need the mapping $\ln\left(\frac{k}{k_*}\right) \rightarrow 
\phi_k$. In the slow roll approximation, this is given by
\begin{equation}
\frac{k}{k_*} = \sqrt{\frac{V}{V_*}} \exp\left\{\int_0^\phi\left|\frac{V}{V_{,\phi}}\right|d\phi\right\}.
\label{kmapping}
\end{equation}

For a given set of parameter values, this integral can be calculated and 
inverted numerically; however, this would be computationally expensive to 
perform while the parameters are being varied during the MCMC procedure 
described in Section \ref{sec:priors}.  Thus, an analytic approximation is 
desirable here, which can be obtained as follows.  Taking the log of equation 
\ref{kmapping}, the formula is easily integrated and inverted if we first 
consider the no-oscillation case where $a=0$.  This yields

\begin{equation}
\phi_{k,0} \equiv \frac{p}{\phi_{min}}\ln\left(\frac{k}{k_*}\right).
\label{phik0}
\end{equation}

For the case where $a \neq 0$, an approximate solution is found by expanding in 
$a \ll 1$ and keeping terms to first order in $\frac{a\phi_{min}}{nf}$. Upon 
integrating, the formula can be inverted approximately by substituting 
equation \ref{phik0} into the sine term, producing the formula

\begin{equation}
\label{phik_approx}
\phi_k \approx \phi_{k,0} - \frac{a \phi_{min}}{p} \left\{\sin\left(\frac{\phi_{k,0}}{f} + \delta\right) - \sin\delta\right\}.
\end{equation}

Using this approximation in eqs.~\ref{powerspec} and \ref{tensor_powerspec}, we 
find the power spectrum differs from the exact numerical solution by less than 
1\% over the range $2 < \ell < 2500$ for all the test cases considered (using 
the approximate formula $\ell \approx k x_c$ where $x_c \approx 14100$ Mpc is 
the comoving angular diameter distance to last scattering).

\section{Sampling and priors}\label{sec:priors}

We sample the model parameter space with a Markov Chain Monte Carlo (MCMC) 
method using the CosmoMC software package (\citealt{lewis2002}), which has been 
modified to incorporate the power spectra in eqs.~\ref{powerspec} and 
\ref{tensor_powerspec} along with the model parameters. For the sampling we use 
the Metropolis-Hastings algorithm extended by a ``fast-slow'' algorithm for 
efficiently sampling nuisance parameters (\citealt{lewis2013}). In addition to 
Planck data (\citealt{plancklike}), the likelihood includes WMAP 9-year 
polarization data (\citealt{hinshaw2013}) as well as small-scale CMB data from 
the ACT (\citealt{story2013}) and SPT surveys (\citealt{sievers2013}).

\subsection{Choice of model parameters}\label{sec:modelparams}

To sample the parameter space, at first it would seem most straightforward to 
vary the five model parameters in equation \ref{potential} ($\phi_{min},p,f,a,$ 
and $\delta$).  However, if this is done then the tensor-to-scalar ratio $r$ 
\emph{cannot} be varied freely, but must be set according to the relation 
$r=16\epsilon_V$, where $\epsilon_V = \frac{1}{2} 
\left(\frac{V_{,\phi}}{V}\right)^2$ is the first potential slow-roll parameter.  
Since $r$ is the more interesting observable, we prefer to vary $r$ freely 
while using this constraint to fix the value of $\phi_{min}$ as a function of 
the other parameters. Substituting the potential from equation \ref{potential}, 
we find our constraint equation,

\begin{equation}
\frac{p}{\phi_{min}}(1-a\sin\delta) - \frac{a}{f}\cos\delta = \sqrt{\frac{r}{8}}.
\label{r_constraint}
\end{equation}

While $a$ gives the amplitude of the oscillation in the potential, the power 
spectrum amplitude is more directly observable in the CMB. By expanding 
equation \ref{powerspec} in the amplitude $a$, it can be seen that the scalar 
power spectrum amplitude is $\approx 2\frac{a\phi_{min}}{nf}$ (this dominates 
over $a$ since $\phi_{min}$ is typically of order 10 while $f$ will be of order 
$0.1$).  Now, if $a$ is exactly zero, then we see from equation 
\ref{r_constraint} that $\frac{p}{\phi_{min}} = \sqrt{\frac{r}{8}}$.  In 
practice, we will find that $a$ must be small, so this will still hold to a 
reasonably good approximation for realistic values of $a$. With this in mind, 
we will define the (approximate) power spectrum oscillation amplitude

\begin{equation}
b \equiv \frac{2a}{f}\sqrt{\frac{8}{r}}.
\label{bdef}
\end{equation}
where, again, typically $b \gg a$. It should be emphasized that our analysis 
will not assume that $b \ll 1$.  While it is true that the power spectrum 
amplitude may differ somewhat from $b$ unless $b \ll 1$, this does not preclude 
our using it as a parameter since the amplitude is still proportional to $b$.  
In practice, our inferred $b$ values will satisfy $b \lesssim 1$, but $b$ will 
not necessarily be very small.

Our power spectrum parameters to vary, then, are $b$, $f$, $\delta$ and $p$, in 
addition to $r$ and the scalar amplitude $A_s$. Expressing equation 
\ref{r_constraint} in terms of $b$, we find

\begin{equation}
\frac{p}{\phi_{min}} = \sqrt{\frac{r}{8}}\left(\frac{1+\frac{1}{2}b\cos\delta}{1-\frac{f}{2}\sqrt{\frac{r}{8}}b\sin\delta}\right).
\label{phimin}
\end{equation}

From the above formula it is obvious that as $b \rightarrow 0$, we recover 
$\frac{p}{\phi_{min}} \approx \sqrt{\frac{r}{8}}$. Equation \ref{phimin} will 
be used to eliminate $\phi_{min}$ in the power spectrum formulae 
(eqs.~\ref{powerspec}, \ref{tensor_powerspec}).

\subsection{e-folding prior}

The number of e-foldings from the time the mode $k_*$ exits the horizon to the 
end of inflation is constrained theoretically to lie within the approximate 
range 50-60.  We will therefore impose a corresponding prior on the number of 
e-foldings, which is given by the integral

\begin{equation}
N = \int_0^{\phi_{e}} \left|\frac{V}{V_{,\phi}}\right|,
\label{efoldings_exact}
\end{equation}
where $\phi_e$ denotes the scalar field value at the end of inflation 
determined by the solution to the equation $\epsilon_V(\phi) \approx 1$. In the 
case with zero amplitude ($a \approx 0$), using the expression for the 
potential in equation \ref{potential} one finds an expression for the number of 
e-foldings $N_0$,

\begin{equation}
\phi_{min}^2 = \frac{p}{2}\left(4 N_0 + p\right).
\label{phimin_efoldings}
\end{equation}
Even for nonzero amplitude, equation \ref{phimin_efoldings} holds approximately 
true, so long as $\phi_{min}$ is defined in terms of the amplitude according to 
equation \ref{phimin}. Combining eqs.~\ref{phimin} and \ref{phimin_efoldings}, 
we find an expression for the approximate number of e-foldings $N_0$:

\begin{equation}
N_0 = p\left\{\frac{4}{r}\left(\frac{1-\frac{f}{2}\sqrt{\frac{r}{8}}b\sin\delta}{1+\frac{1}{2}b\cos\delta}\right)^2 - \frac{1}{4}\right\}.
\label{efoldings_formula}
\end{equation}

To determine the exact number of e-foldings $N$, the value for $\phi_{e}$ and 
the integral in equation \ref{efoldings_exact} must be calculated numerically.  
In practice, the presence of the oscillations cause $N$ to differ from that 
determined by equation \ref{phimin_efoldings} only by a small amount (typically 
$N>N_0$ by less than 3), although it depends on the oscillation amplitude and 
period.  Since the integral in equation \ref{efoldings_exact} is 
computationally expensive to calculate while varying parameters, we will 
instead use equation \ref{efoldings_formula} for the approximate number of 
e-foldings to enforce a prior in $N_0$ during the MCMC routine.

While equation \ref{efoldings_formula} typically gives a reasonable 
approximation to the number of e-foldings, a catastrophe can occur near the end 
of inflation if the oscillations dominate over the monomial term in the 
potential---in this case the potential may cease to become monotonic and a 
local minimum (false vacuum) can occur, rendering the number of e-foldings 
effectively infinite.\footnote{To be precise, a very large number of e-foldings 
would occur until the field tunnels out of the local minimum via bubble 
nucleation. Given that relatively few e-foldings would follow this before 
inflation ends, this scenario would produce an open universe with a very 
sub-critical energy density, in gross violation of cosmological constraints 
(see for example \citealt{bucher1995}).} This occurs if either the amplitude 
$b$ or the exponent $p$ are too large; in the latter case, the slope of the 
monomial term becomes very shallow before inflation ends, allowing the 
oscillations to dominate.  To deal with this issue, we will refine the 
e-folding prior during post-processing by performing the numerical integral 
(equation \ref{efoldings_exact}) to find $N$ for each point in the MCMC chain.  
This allows us to eliminate regions of parameter space where the number of 
e-foldings $N$ becomes large or infinite.

We therefore obtain a flat prior in the number of e-foldings as follows. During 
the MCMC routine, we sample the parameter space with a flat prior in the 
approximate number of e-foldings $N_0$ over a liberal range from 40 to 70.  
During post-processing, the exact number of e-foldings $N$ for each point in 
parameter space is calculated by first finding the field value at the end of 
inflation $\phi_e$ via a grid search. If a local stationary point is 
encountered before inflation ends, then the number of e-foldings is effectively 
infinite and thus the point is discarded. For the remaining points, we compute 
$N$ by performing the integral in equation \ref{efoldings_exact} numerically; 
points with $N$ outside the canonical range from 50 to 60 are then discarded.  
By this method, we obtain a flat prior in the number of e-foldings $N$ to good 
approximation (this will be verified in Section \ref{sec:results}).

One last subtlety remains in implementing the e-folding prior: the parameter 
$N_0$ is entirely determined by the model parameters discussed in Section 
\ref{sec:modelparams} and thus cannot be included as a separate parameter. We 
therefore impose the $N_0$ prior by making a transformation of variables. In 
Section \ref{sec:results} we will show that $\delta$ prefers to be small, and 
since $f\sqrt{\frac{r}{8}} \ll 1$, we can approximate equation 
\ref{efoldings_formula} as

\begin{equation}
N_0 \approx \frac{4p}{r}\left(1+\frac{1}{2}b\right)^{-2}.
\label{efoldings_formula_approx}
\end{equation}
Thus, to a good approximation the number of e-foldings depends only on the 
parameters $p$, $r$, and $b$, with very little dependence on $f$ and $\delta$.  
Since $N_0$ will not be one of our primary model parameters, we enforce the 
e-folding prior by starting with $N_0$ as a parameter (with a given prior), 
then making the transformation from $N_0$ to $b$.  We then derive our prior in 
$b$ from the prior in $N_0$ and the resulting Jacobian $|\partial N_0/\partial 
b|$ using equation \ref{efoldings_formula}. As we will see in Section 
\ref{sec:results}, $b$ is fairly well-constrained (apart from a small 
non-Gaussian tail) and the Jacobian has only a minor effect on the inferred 
parameters.

\subsection{Priors in the model parameters}

Apart from the amplitude $b$ whose prior is defined by the e-folding constraint 
(equation \ref{efoldings_formula}), we choose a flat prior in the remaining 
model parameters ($p$, $r$, $f$, $\delta$). Here we discuss the preferred range 
of each parameter.

In order to sample the full range of possible phases and amplitudes, 
the most straightforward approach would be to vary $\delta$ over the range 
($-\pi,\pi$) and allow $b$ to vary from 0 to some large amplitude $b_{max}$.  
However, the same can be achieved by varying $\delta$ in the range 
($-\frac{\pi}{2},\frac{\pi}{2}$) and allowing $b$ to have a \emph{negative} 
amplitude, since the point ($b$,$\delta=\pi$) is equivalent to 
(-$b$,$\delta=0$). We favor the latter approach, since it avoids having a 
bimodal posterior distribution in the phase shift $\delta$. A negative 
amplitude corresponds to having a \emph{positive} running of the spectral index 
near the pivot scale, and the posterior will run continuously from positive to 
negative $b$-values. We choose a liberal $b_{max}=2$ so our allowed range in 
$b$ will therefore be $(-2,2)$.

We can get a sense of the desired range in $f$ by considering the oscillation 
period in $\log \ell \sim \log k + const$. Combining eqs.~\ref{powerspec} and 
\ref{phik0}, we find that the axion decay constant $f$ is related to the period 
in $\log k$ by

\begin{equation}
f = \frac{\ln 10}{2\pi}\left(\frac{p}{\phi_{min}}\right)P_{\log k}
\label{Plogk_exact}
\end{equation}
Using equation \ref{phimin_efoldings}, this becomes
\begin{equation}
f \approx 0.26 P_{\log k} \sqrt{\frac{p}{N_0}}.
\label{Plogk}
\end{equation}

From the largest scales down to the Silk damping tail probed by ACT/SPT, 
observable primary anisotropies in the CMB span a range of $\ell \approx 
2-3500$, corresponding to $\Delta \log \ell \approx 3$. At higher multipoles, 
there is no strong preference for negative running in either the Planck or ACT 
data (although SPT does prefer negative running at high $\ell$ to some 
extent---see \citealt{valentino2010}).  We therefore expect at worst a mild 
suppression of power at high $\ell$. In order to have suppression at low $\ell$ 
without an equally large suppression at higher $\ell$, the entire range of 
multipoles $\Delta\log \ell\sim 3$ should fit less than roughly three quarters 
of a full period; in other words, the period $P_{\log \ell} = P_{\log k}$ 
should be roughly greater than $\sim 4$.  For $p$ running from $0.5$ to $4$ and 
50-60 e-foldings, we find from equation \ref{Plogk} that $f$ should be larger 
than $\approx 0.1$.

On the other hand, a large period of oscillation would require a large 
amplitude to achieve the desired suppression of power at low $\ell$, and this 
would approach the constant running limit (over CMB scales) that has been 
considered before.  For $f\approx 1$, the oscillation period is at least 
several times larger than the relevant $\ell$-range (unless $p \ll 1$) such 
that the running is effectively constant in this regime. As we will see in 
Section \ref{sec:results} however, regions of parameter space with a large 
period $P_{\log k}$ are forbidden by the e-folding constraint unless the 
amplitude $b \lesssim 1$, and hence the running of the spectral index is small 
(see also appendix \ref{sec:3dposts}).  Thus, allowed solutions with $f \gtrsim 
1$ tend to have small running and cannot fit the low-$\ell$ power spectrum 
significantly better than the usual power-law power spectrum model. In Section 
\ref{sec:results} we will show that this expectation is correct and thus the 
constraints do not change significantly when $f$ is extended up to 2 (see 
Figure \ref{fig:fpriors}).  We therefore choose the fiducial range to be $f \in 
(0.1,1)$.

Given that the period $P_{\log k}$ is more directly observable in the power 
spectrum than $f$ is, it might be tempting to use $P_{\log k}$ as our model 
parameter instead of $f$. We choose not to do this, for two reasons.  On the 
theory side, a super-Planckian axion decay constant $f \gtrsim 1$ is difficult 
to implement in the underlying string theory; in fact $f \ll 1$ seems necessary 
to embed the corresponding model, although it may be possible for multiple 
axions to combine to produce a larger effective axion decay constant 
(\citealt{kim2005}; \citealt{kappl2014}).  Given these difficulties, we prefer 
to impose an upper limit $f \leq f_{max}$, with $f_{max}=1$ being the fiducial 
value.  This does \emph{not} translate to a clean upper bound in $P_{\log k}$, 
however, since from equation \ref{Plogk} it is possible to have $f \gtrsim 1$ 
even if $P_{\log k}$ is relatively small provided $p$ is large enough.

The second reason for choosing $f$ as our parameter instead of $P_{\log k}$ is 
more subtle.  As we will see in Section \ref{sec:results}, the posterior 
distribution is multi-modal and there exist regions of parameter space which 
fit the high-$\ell$ likelihood at the expense of lower $\ell$. For example, if 
$p$ is made small, by equation \ref{Plogk} the period becomes large unless $f$ 
is also small.  However in the latter case, we find another mode emerges which 
has negligible running and only fits the high-$\ell$ likelihood (for discussion 
see appendix \ref{sec:3dposts}).  Since we are primarily interested in 
improving the fit at low as well as high multipoles, by imposing a lower bound 
$f_{min}=0.1$ we avoid this spurious mode entirely. Thus, our fiducial range in 
$f$ is ($0.1,1$). In Section \ref{sec:nr_constraints} we will consider the 
effect of varying the prior of $f$, including the allowed range (see Figure 
\ref{fig:fpriors}).

\begin{figure*}[t]
	\centering
	\includegraphics[height=1.0\hsize,width=1.0\hsize]{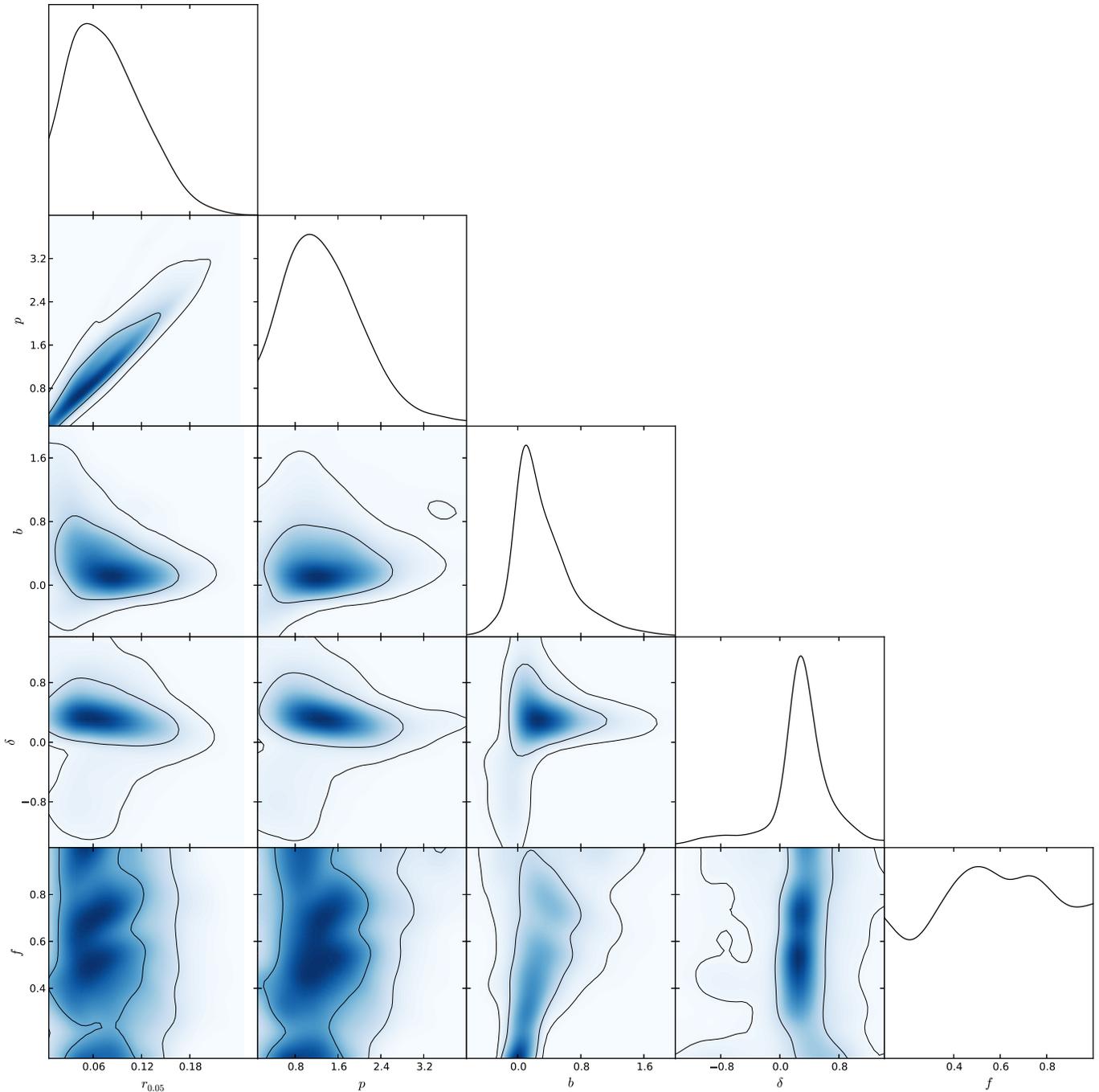}
\caption{Posteriors in the inflation model parameters ($p$, $b$, $\delta$, $f$) 
discussed in Section \ref{sec:modelparams} and the tensor-to-scalar ratio $r$ 
at the pivot scale $k_*=0.05$ Mpc$^{-1}$.  Contours enclose 68\% and 95\% of 
the total probability in each joint posterior.  Note the small probability for 
$r=0.2$, which lies just outside the 99\% confidence region.  The correlation 
between the exponent $p$ and $r$ is a consequence of the e-folding constraint 
(eq.~\ref{efoldings_formula}). In contrast to the amplitude $b$ and phase shift 
$\delta$ which have peaked distributions, the axion decay constant $f$ is 
poorly constrained although it exhibits a mild correlation with $b$. Finally, 
note that the distributions in $r$ and $p$ exhibit very little dependence on 
$f$, and are thus fairly insensitive to the assumed prior in $f$.}
\label{fig:triangle_plot}
\end{figure*}

\section{Results}\label{sec:results}

\subsection{Constraints on the oscillation parameters $b$, $\delta$, 
$f$}\label{sec:axion_constraints}

After sampling the parameter space with the data and priors discussed in 
Section \ref{sec:priors}, we display marginal posterior probability 
distributions in the model parameters in a ``triangle plot'' in Figure 
\ref{fig:triangle_plot}.  Starting with the one-dimensional posteriors, it is 
clear that the oscillation amplitude $b$ and phase shift $\delta$ have 
distributions that are well peaked, albeit with significant non-Gaussian tails.  
By contrast, the axion decay constant $f$ is very poorly constrained, with a 
mild preference for $f \approx 0.5$ but largely prior-dominated.

\begin{figure*}[t]
	\centering
	\subfigure[spectral index]
	{
		\includegraphics[height=0.30\hsize,width=0.38\hsize]{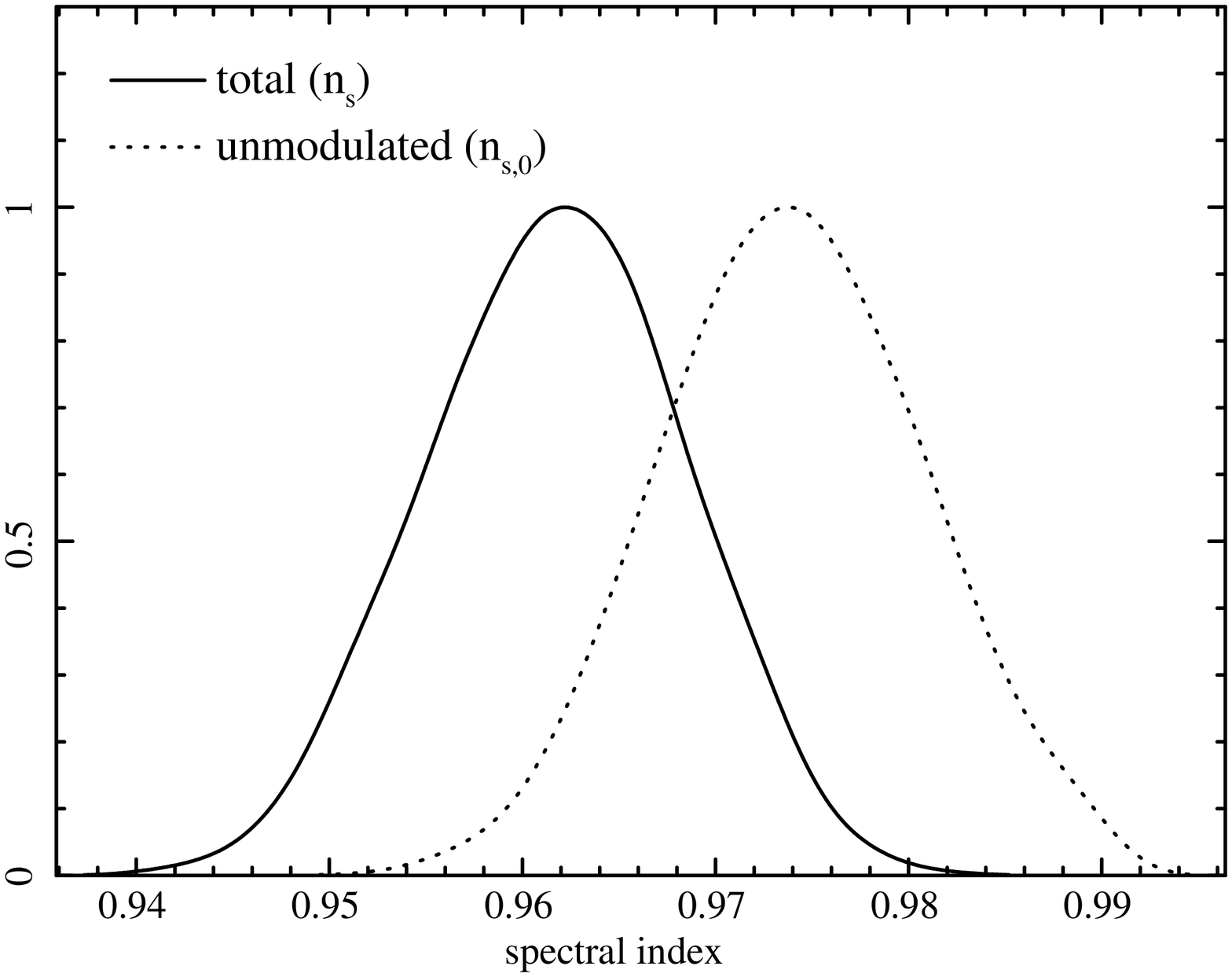}
		\label{post_ns}
	}
	\subfigure[running of the spectral index $\alpha_*$]
	{
		\includegraphics[height=0.30\hsize,width=0.38\hsize]{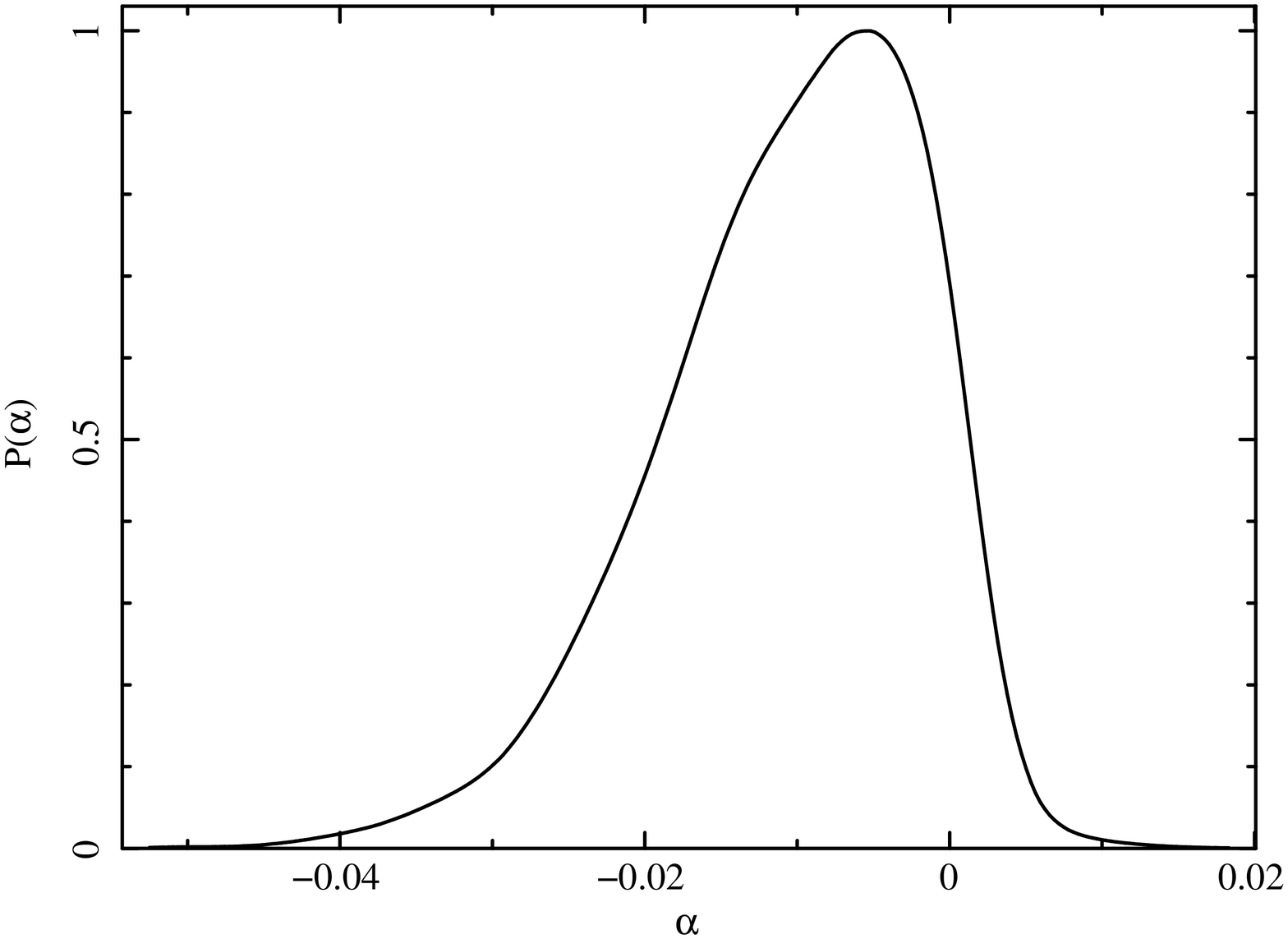}
		\label{post_alpha}
	}
	\subfigure[number of e-foldings $N$]
	{
		\includegraphics[height=0.30\hsize,width=0.38\hsize]{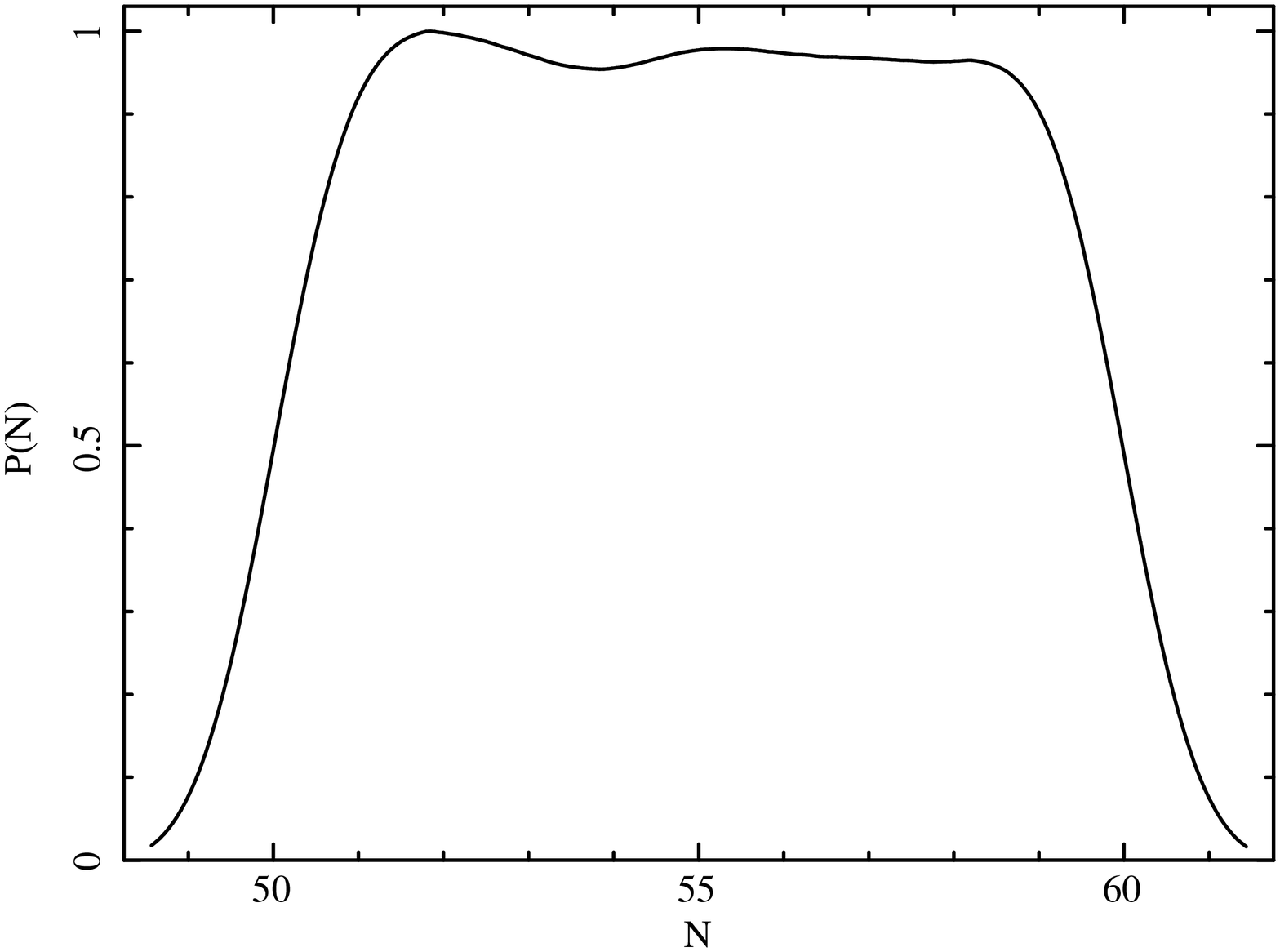}
		\label{post_N}
	}
	\subfigure[oscillation period $P_{\log k}$]
	{
		\includegraphics[height=0.30\hsize,width=0.38\hsize]{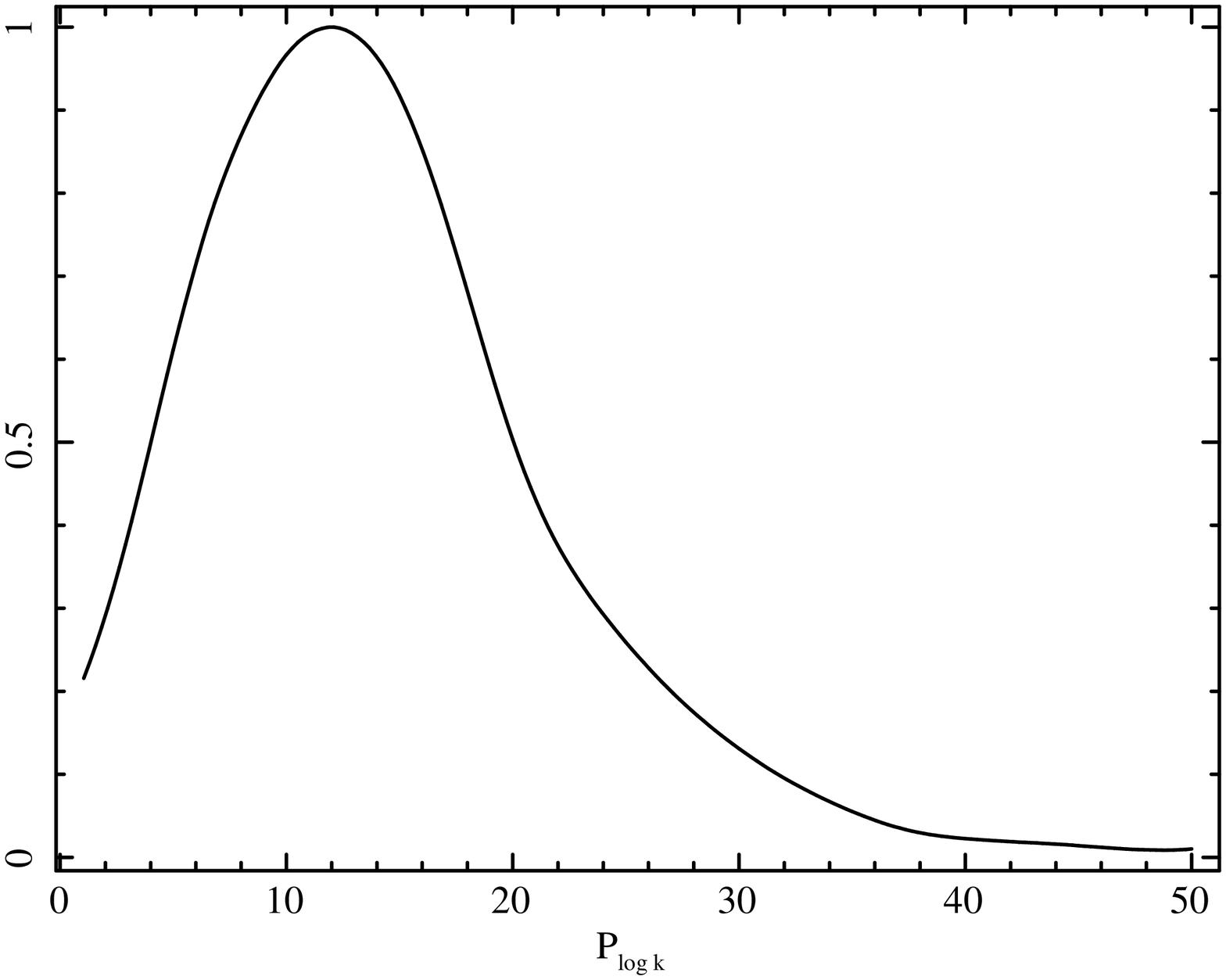}
		\label{post_Plogk}
	}
	\caption{Marginal posterior probability in four derived parameters: (a) The 
spectral index at the pivot scale $k_* = 10$ Mpc$^{-1}$. The total spectral 
index $n_s$ (solid line) is plotted together with the unmodulated spectral 
index $n_{s,0}$ (dashed line) which excludes the oscillatory part. (b) Running 
of the spectral index $\alpha_*$ at the pivot scale. (c) The number of 
e-foldings $N$ from the time that mode $k_*$ left the horizon, to the end of 
inflation.  (d) Period of oscillation in the power spectrum in terms of 
$\log_{10}k$.}
	\vspace{10pt}
\label{fig:post_derivedparams}
\end{figure*}

Before proceeding, is important to verify that the constraints on the spectral 
index $n_s$ and running $\alpha_*$ at the pivot scale $k_*=0.05$ Mpc$^{-1}$ are 
consistent with those obtained from the usual power-law spectrum and constant 
running models. To this end, analytic expressions for $n_s$ and $\alpha_*$ are 
derived in Appendix \ref{sec:ns_alpha_formulas} and given in equations 
\ref{ns_formula} and \ref{alpha_formula}, respectively. Using these formulas we 
plot the corresponding derived posteriors in Figures \ref{post_ns} and 
\ref{post_alpha}. The probability is peaked around $n_s\approx 0.96$, entirely 
consistent with the base Planck model; likewise, the allowed running $\alpha_*$ 
lies in the approximate range $(-0.03,0)$, consistent with the usual constant 
running model. In Figure \ref{post_N} we plot a posterior in the number of 
e-foldings $N$ (calculated from eq.~\ref{efoldings_exact}), which shows $N$ to 
be entirely dominated by the chosen flat prior in the fiducial range 50-60.  
Finally, in Figure \ref{post_Plogk} we plot a posterior in the oscillation 
period $P_{\log k}$ (using eq.~\ref{Plogk_exact}), which is well-peaked in 
contrast to the constraint on $f$.

Returning to the model parameter constraints, the structure of the posterior 
distribution can be better understood from the joint 2-dimensional posteriors 
in Figure \ref{fig:triangle_plot}. In the $f$ vs. $b$ plot, one sees a mild 
correlation in $f$ and $b$, particularly for $f \lesssim 0.6$. This can be 
understood in terms of the running of the spectral index: if the period 
(corresponding to $f$) is increased, one must also increase the amplitude $b$ 
to keep the running $\alpha_*$ constant (this can also be seen in the formula 
for $\alpha_*$ in equation \ref{alpha_formula}).  The correlation is not very 
tight, however, because the required running depends on $r$: the greater the 
tensor contribution at low $\ell$, the greater the (negative) running must be 
to adequately suppress a corresponding amount of low-$\ell$ scalar power.

In Figure \ref{fig:triangle_plot}, we can see in the $b$ vs.  $\delta$ plot the 
distribution is obviously multi-modal, where outside the maximum probability 
region there are three separate ``wings''. Two of the wings run off to large or 
small $\delta$, while the amplitude $b$ is very small or even negative. The 
third wing runs off to high amplitudes $b$.  The multi-modal structure is 
discussed in further detail in appendix \ref{sec:3dposts}; here we will simply 
note that in each of these wings, $f$ takes on either very small or large 
values (as expected since $b$ and $f$ are correlated). If one stays within the 
high probability region, $f$ is somewhat better constrained than the 
$f$-posterior in Figure \ref{fig:triangle_plot} would suggest.

\subsection{Best-fit model}\label{sec:bestfit}

Next, to find the best-fit point by minimizing the likelihood requires some 
care, because the exact number of e-foldings $N$ is computationally too 
expensive to perform during the minimization procedure. We therefore obtain the 
best-fit point by factoring in a prior in $N_0$ (given by 
eq.~\ref{efoldings_formula}), where in this case our prior is chosen to be 
strongly peaked (a Gaussian with dispersion $\sigma_{N_0}$=0.5) about a 
particular value $N_{0,i}$.  This procedure is performed over a grid of 16 
values $N_{0,i}$ regularly spaced from 45 to 65.  After finding each best-fit 
point, the exact number of e-foldings $N$ is calculated for each; the best-fit 
point with the highest likelihood value that remains within the range $N \in 
[50,60]$ is chosen as the global best-fit point.

The resulting best-fit parameter values are given in the first row of Table 
\ref{tab:bestfit}. Errors are given for the inflation parameters $r$, $p$, $b$ 
and $\delta$, determined by the 16\% and 84\% percentiles of the posterior 
probability distribution in each parameter.

\begin{figure}
	\includegraphics[height=0.7\hsize,width=0.92\hsize]{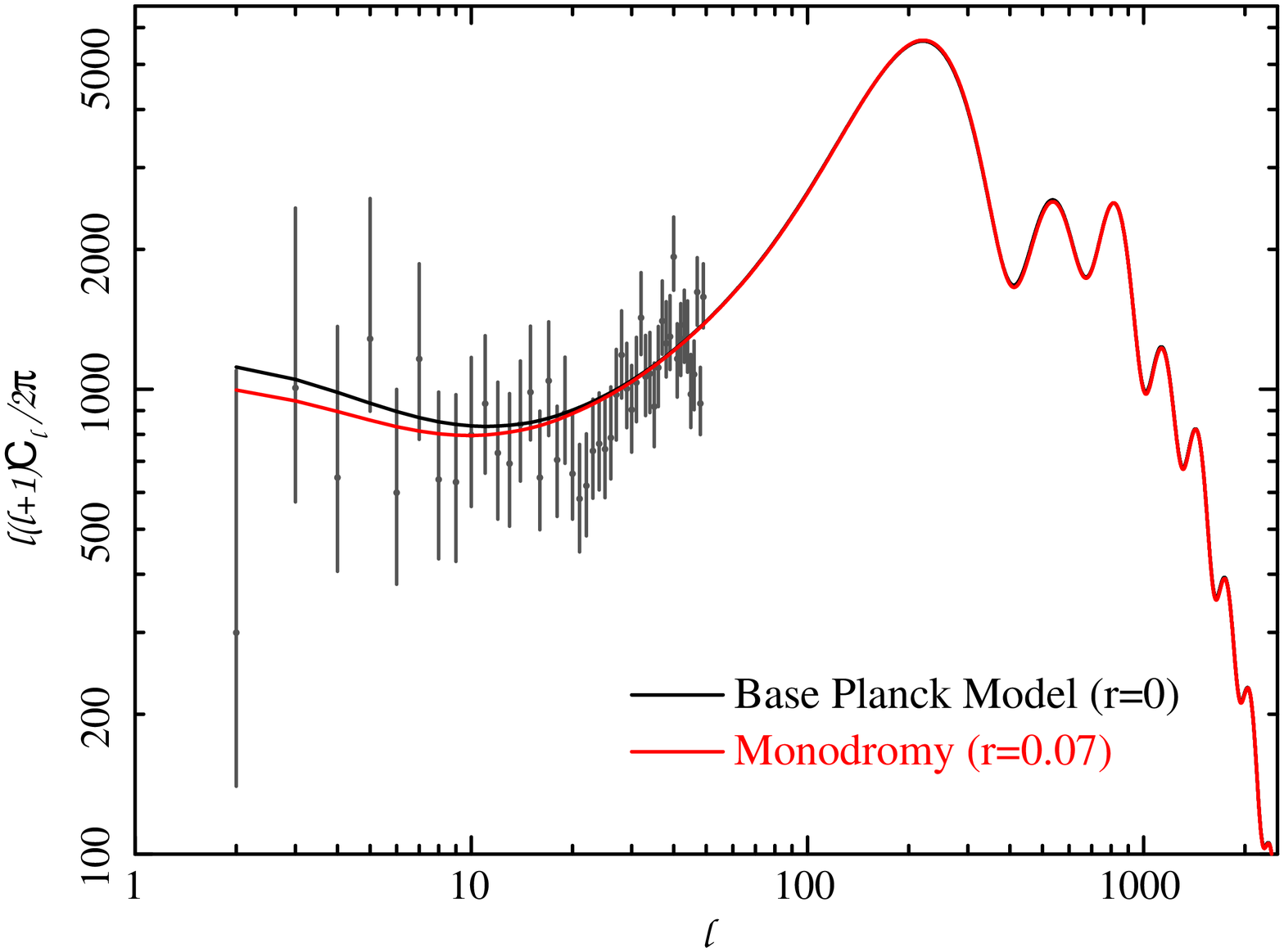}
\caption{Best-fit TT angular power spectrum for the base Planck model with zero 
running (dark line), and axion monodromy model (red line). Error bars are shown 
for $2 \leq \ell < 50$. These best-fit spectra are determined using a 
combination of Planck+lensing+WP+high $\ell$ data.}
\label{fig:allcls}
\end{figure}

In Figure \ref{fig:allcls} we plot the CMB angular power spectrum over the 
multipole range $2 \leq \ell \leq 2500$ for the best-fit axion monodromy model 
(dark line) compared to the base Planck model (red line; defined as the 
$\Lambda$CDM model with a power-law spectrum, i.e.~zero running).  Note that 
for $\ell \gtrsim 30$, the two models are indistinguishable, while at lower 
multipoles the axion monodromy power is suppressed by up to $\approx$ 20\% 
compared to the base Planck model.  This suppression is a consequence of the 
running spectral index coming from the sinusoidal term in the potential 
(eq.~\ref{potential}).

\begin{table*}
\centering
\begin{tabular}{|l|c|c|c|c|c|c|c|c|c|c|c|c|c|c|}
\hline
& $r$ & $p$ & $b$ & $\delta$ & $f$ & $10^9 A_s$ & $H_0$ & $\Omega_m$ & $\Omega_b$ & $\tau$ & $N$ & $\alpha_*$ & $n_s$ & $n_{s,0}$\\
\hline
Best-fit & $0.07^{+0.05}_{-0.04}$ & $1.55^{+0.56}_{-0.92}$ & $0.44^{+0.24}_{-0.45}$ & $0.30^{+0.32}_{-0.31}$ & 0.53 & 2.209 & 67.9 & 0.307 & 0.048 & 0.093 & 58.5 & -0.014 & 0.959 & 0.979\\
\hline
$r=0.13$ & 0.13 & 2.37 & 0.41 & 0.20 & 0.58 & 2.216 & 68.2 & 0.304 & 0.048 & 0.094 & 59.9 & -0.020 & 0.960 & 0.970\\
$r=0.19$ & 0.19 & 3.08 & 0.31 & 0.12 & 0.54 & 2.210 & 68.3 & 0.303 & 0.048 & 0.093 & 51.9 & -0.025 & 0.960 & 0.961\\
\hline
\end{tabular}
\caption{Best-fit axion monodromy models}
\label{tab:bestfit}
\end{table*}

\subsection{Constraints on the tensor-to-scalar ratio $r$ and potential 
parameter $p$}\label{sec:nr_constraints}

Marginal posteriors in $p$ and $r$ are shown in Figure \ref{fig:triangle_plot}.  
Strikingly, the BICEP2 result $r = 0.2$ lies just outside the 99\% confidence 
region, while the highest probability $r$-value is $\approx 0.06$, similar to 
the best-fit value $r \approx 0.07$ (Table \ref{tab:bestfit}). Likewise, $p > 
3$ lies outside the 95\% confidence region.

From the e-folding constraint given by equation \ref{efoldings_formula} it is 
apparent that the exponent $p$ and the tensor-to-scalar ratio $r$ should be 
strongly correlated, provided the amplitude $b$ does not get too large. The 
joint posterior in $p$ and $r$ shown in Figure \ref{fig:triangle_plot} shows 
that this is indeed the case; a high $r$-value is correlated with a large 
exponent $p$.  The width of the posterior around this correlation is determined 
by the allowed number of e-foldings, and also the range of $b$-values.

\begin{figure}
	\includegraphics[height=0.7\hsize,width=0.92\hsize]{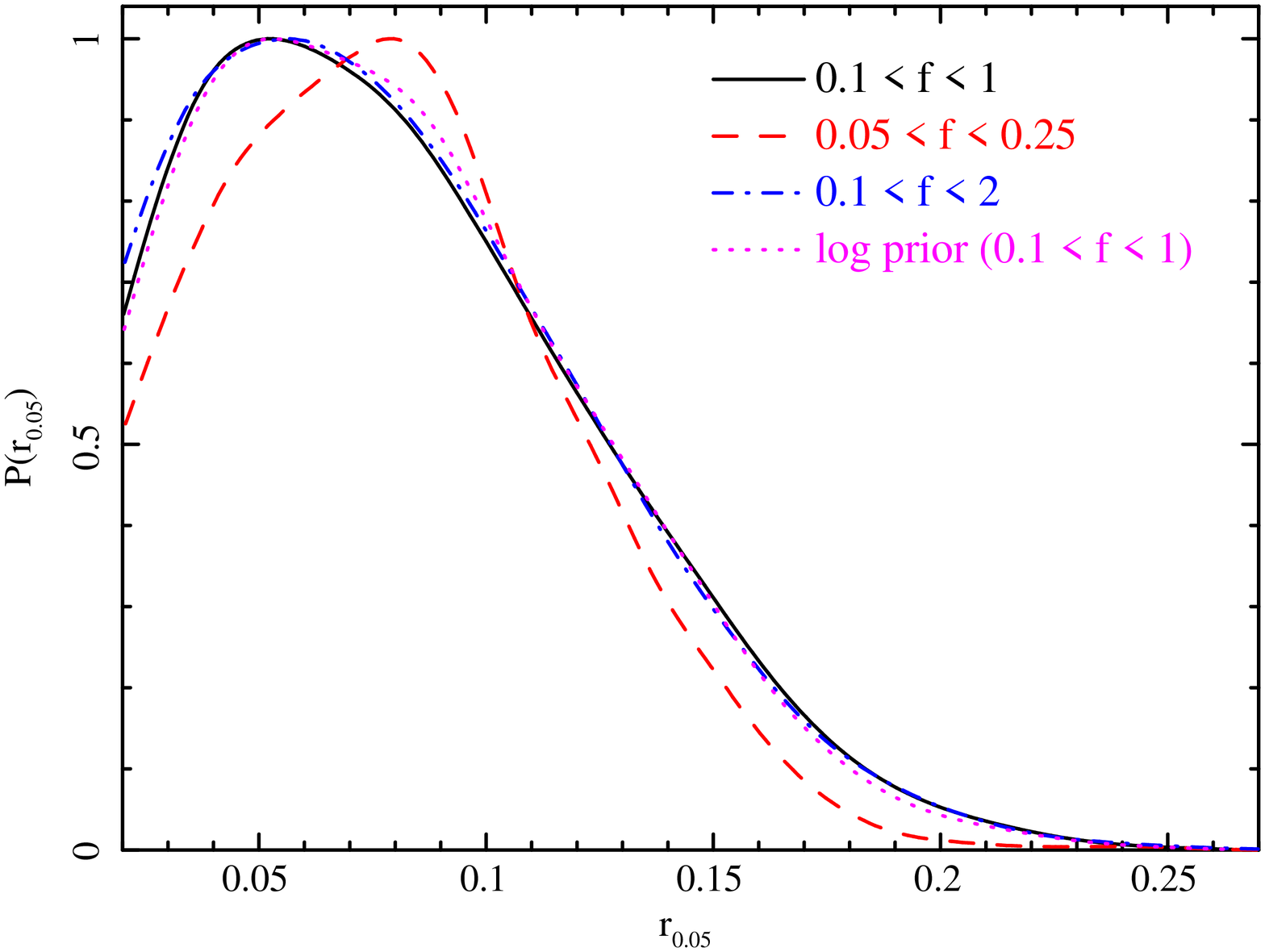}
\caption{Posterior distribution in the tensor-to-scalar ratio $r$ for four 
different priors in $f$: three uniform priors covering different ranges in $f$ 
(solid, dashed, dot-dashed lines) and a log prior in $f$ (dotted line). The 
inferred probability in $r$ (and likewise $p$) is quite robust since it is 
insensitive to the assumed prior in $f$, while the other model parameters 
($b$,$\delta$) are fairly well-constrained.}
\label{fig:fpriors}
\end{figure}

Since the axion decay constant $f$ is poorly constrained, it is important to 
check whether these results are sensitive to the chosen prior on $f$. From the 
$r$ vs. $f$ and $p$ vs. $f$ plot in Figure \ref{fig:triangle_plot}, one can 
anticipate that the inferred probability of $r$ and $p$ are weakly dependent on 
$f$, since essentially no correlation with $f$ is seen for either parameter. To 
verify this, in Figure \ref{fig:fpriors} we plot posterior distributions in $r$ 
for four different assumed priors in $f$: a flat prior in the fiducial range $f 
\in (0.1,1)$ (solid line), a flat prior in the ranges $f \in (0.05,0.25)$ (dashed 
line) and $f \in (0.1,2)$ (dot-dashed line), and finally a log prior in $f$ 
over the fiducial range. As can be seen in the figure, the inferred probability 
in $r$ (and likewise $p$) is quite robust to changes in the assumed $f$ prior.  
Since the other model parameters ($b$,$\delta$) are fairly well-constrained, 
our result for $r$ is thus quite insensitive to the priors in the axion model 
parameters.

\begin{figure}
	\includegraphics[height=0.9\hsize,width=0.92\hsize]{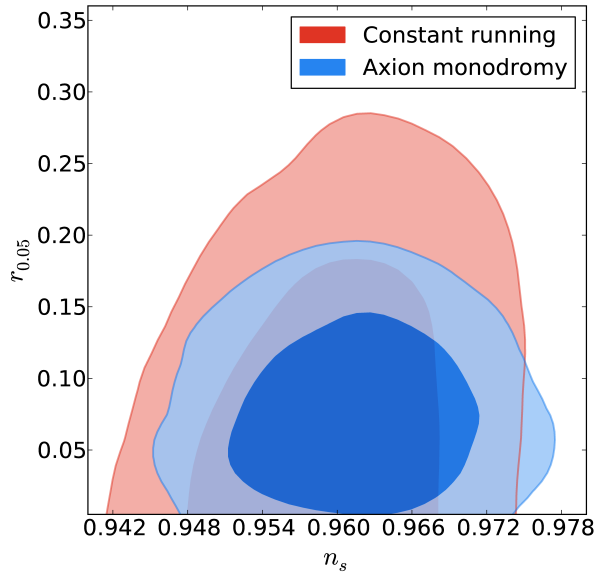}
\caption{Joint posteriors in the spectral index $n_s$ and the tensor-to-scalar 
ratio $r$, both evaluated at the pivot scale $k=0.05$ Mpc$^{-1}$.  Constraints 
(68\% and 95\%) are shown for the constant running model (red) and axion 
monodromy model (blue). While the constant running model prefers $r=0$, in 
axion monodromy the highest probability occurs for $r\approx 0.07$ because a 
nonzero $r$-value is required to fit the running and spectral index well.  Note 
also the constant running model allows for a higher $r$, primarily because it 
is not subject to the e-folding constraint.}
\label{fig:r_vs_ns}
\end{figure}

Why does the data prefer a small (but nonzero) $r$ value? Generally, higher $r$ 
means a larger tensor contribution to $TT$ anisotropies at low $\ell$. This 
exacerbates the deficit in power at low $\ell$, and thus requires a higher 
negative running to make up for it. However, the likelihood at high $\ell$ 
shows no strong preference for negative running (\citealt{planck2013}), and the 
fit worsens at high $\ell$ as the running increases. Thus, the fit generally 
worsens as $r$ is increased toward large values. This is not the whole story 
however, because $r$ is significantly more constrained in the axion monodromy 
model compared to the usual constant running model, for which $r < 0.26$ at the 
95\% confidence level (\citealt{planck2013}). This can be seen in Figure 
\ref{fig:r_vs_ns}, where we plot a joint posterior in $r$ vs. $n_s$ for the 
constant running model (red) and the axion monodromy model (blue). While both 
models prefer a similar $n_s$, $r$ is more constrained in axion monodromy: the 
point $r=0.2$, $n_s\approx 0.96$ lies outside the 95\% CL contour for axion 
monodromy, while it is well within the same contour for the constant running 
model.

Additionally, note from Figure \ref{fig:r_vs_ns} that $r$ prefers to be zero in 
the constant running model, whereas the axion monodromy model has its peak 
probability near the best-fit $r\approx 0.07$. The reason why axion monodromy 
does not prefer $r=0$ is simple: one can see from equations \ref{ns_formula} 
and \ref{alpha_formula} that $n_s$ and $\alpha_*$ are dependent on $r$, whereas 
in the constant running model, these parameters can be chosen independently of 
$r$.  The best-fit constant running model has a zero tensor contribution 
($r=0$) and running $\alpha \approx -0.012$. This shows that having a nonzero 
tensor contribution cannot be entirely made up for by negative running---even 
if running is allowed, the likelihood itself still prefers $r=0$.  In the axion 
monodromy model however, this solution is impossible, since the running is 
proportional to $r$ according to equation \ref{alpha_approx}. Thus, $r\approx 
0$ would necessarily imply negligible running, and likewise the spectral index 
would be a poor fit. Instead, the best-fit axion monodromy model settles for an 
$r$-value ($\approx 0.07$) which is high enough to give the necessary running 
and spectral index, but low enough that the tensor contribution doesn't spoil 
the fit too much.  This compromise necessarily results in a slightly worse fit 
at low $\ell$ compared to the constant running model.

We still need to understand why axion monodromy is more restrictive at the 
large $r$ end compared to constant running. To investigate this, Table 
\ref{tab:delta_chisq} shows the change in $\chi^2$ for four models compared to 
the base Planck model with zero running ($\alpha=0$).  The four models are: the 
best-fit axion monodromy model (for which $r \approx 0.07$), the best-fit 
constant running model (for which $r=0$), and the same two best-fit models when 
$r$ is fixed to 0.19 rather than varied.  Note that while the total $\chi^2$ is 
decreased by a similar amount in both best-fit models, $\chi^2$ is increased in 
the $r=0.19$ models, with the axion monodromy $r=0.19$ giving the worst fit. To 
understand why this is the case, we must break this down into the various 
likelihoods representing different multipole ranges.

Starting with the high-$\ell$ likelihoods, note that while the fit to the 
CAMspec likelihood (comprising the majority of multipoles in the Planck data) 
is barely affected for the general best-fit models, the fit is dramatically 
worsened in the models for which $r=0.19$. This is a consequence of the 
aforementioned high running required to fit the low-$\ell$ likelihood well for 
large $r$. The same is true (though not as dramatically) for the ACT/SPT 
likelihoods. For the Commander likelihood (low $\ell$), on the other hand, all 
models show an improved fit, but the constant running model gives a 
significantly better fit; this disparity is greater for the $r=0.19$ models. In 
the following section we investigate the nature of this disparity more closely 
and why it worsens with large $r$.

\subsection{Comparison to constant running model at low multipoles}\label{sec:lowl}

\begin{figure}
	\includegraphics[height=0.7\hsize,width=0.92\hsize]{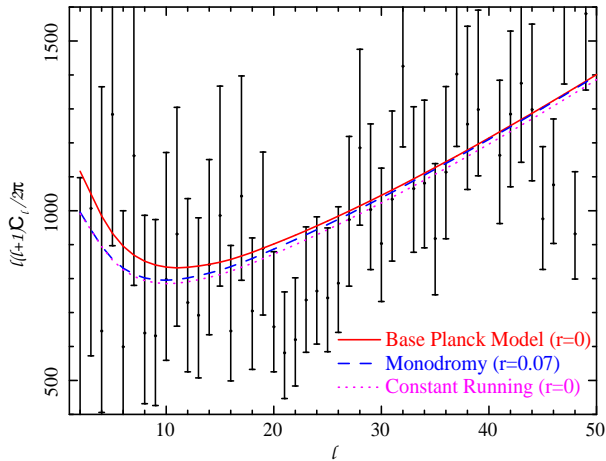}
\caption{Best-fit CMB $TT$ angular power spectrum at low multipoles for the 
base Planck model with zero running (red solid line), the axion monodromy model 
(blue dashed line), and the constant running model (magenta dotted line).  
Although axion monodromy and the constant running model exhibit a similar 
suppression of power at the lowest multipoles, the latter has a greater 
suppression of power in the range $10 < \ell < 50$, partly because there is no 
tensor contribution ($r=0$) and because it is not subject to the e-folding 
constraint.}
\label{fig:lowcls}
\end{figure}

In Figure \ref{fig:lowcls} we plot the best-fit angular $TT$ power spectrum for 
$2 \lesssim \ell \lesssim 50$ for the base Planck model with zero running (red 
solid line), axion monodromy (blue dashed line), and constant running (magenta 
dotted line) models. The black error bars give the errors in each individual 
multipole. Note that while both models achieve similar reduction in power at 
very low $\ell$, the axion monodromy model achieves very little reduction in 
power for $\ell \gtrsim 25$. The deficit in power in the data (points with 
error bars) is most noticeable over the range $20 \lesssim \ell \lesssim 30$, 
and in this range the constant running model is a significantly better fit.

Why does the constant running model fit the low-$\ell$ likelihood better? There 
are a few reasons. As discussed in the previous section, axion monodromy does 
not have the freedom to choose $r$ and $\alpha$, $n_s$ independently. In 
particular, $r=0$ is disfavored because it produces a negligibly small running 
and incorrect spectral index. Unfortunately however, a nonzero $r$ slightly 
worsens the fit at low $\ell$ by including a tensor contribution.

\begin{figure}
	\includegraphics[height=0.7\hsize,width=0.92\hsize]{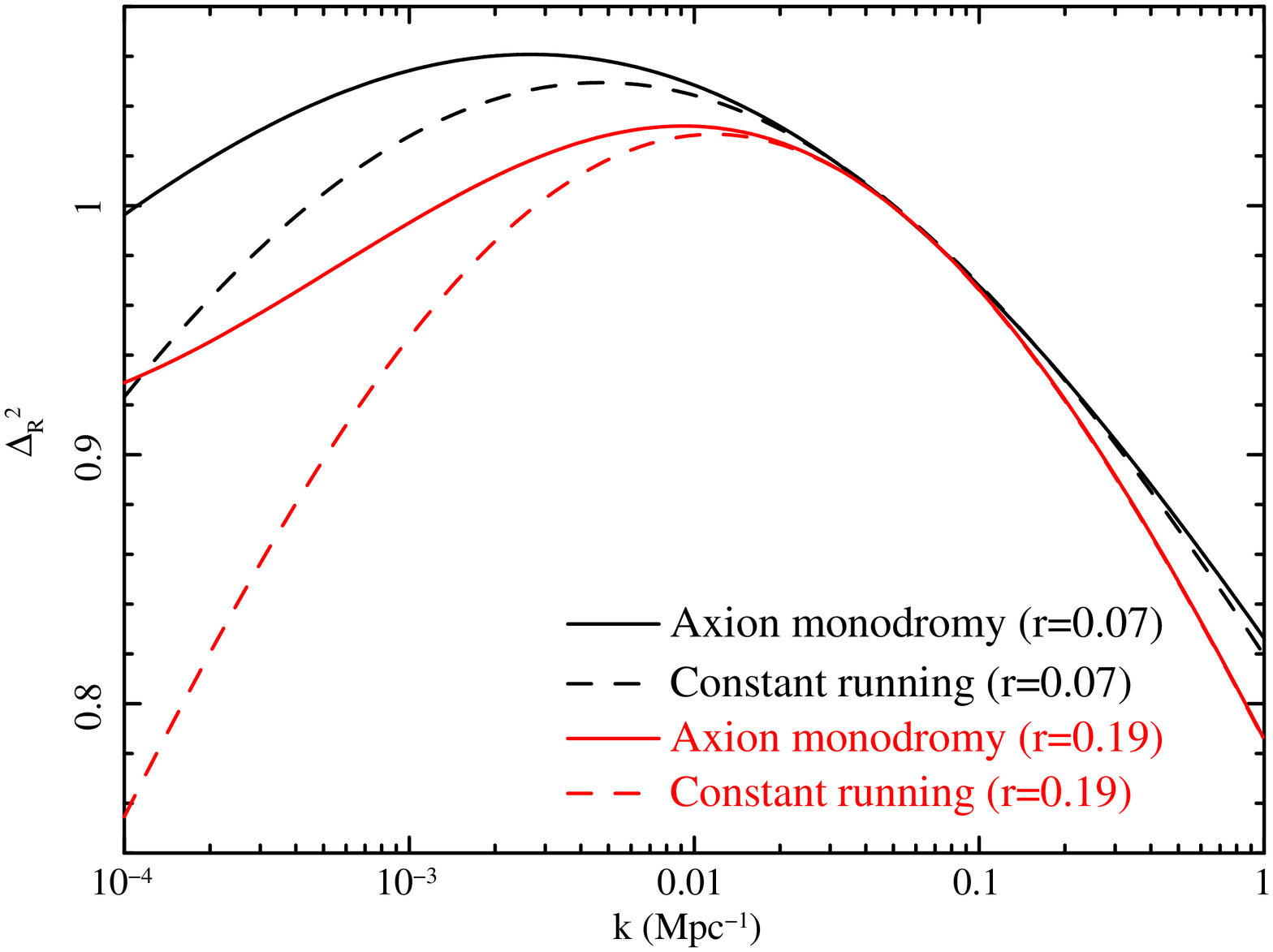}
\caption{Best-fit scalar power spectrum for axion monodromy (solid lines) vs.  
constant running models (dashed lines), all normalized to 1 at $k=0.05$ 
Mpc$^{-1}$ for the sake of comparison.  For $r=0.07$ (dark lines), axion 
monodromy achieves only slightly less suppression at low $k$ (corresponding to 
low $\ell$) compared to the constant running model.  For $r=0.19$ (red lines) 
however, axion monodromy has milder negative running at low $k$ and thus cannot 
achieve enough suppression to make up for the tensor contribution at low 
$\ell$.  This is a consequence of the e-folding constraint, which enforces a 
large exponent $p$ and limits the oscillation amplitude $b$ of the power 
spectrum.}
\label{fig:axion_vs_run}
\end{figure}

Even if $r$ is fixed to the same value in both models, however, the constant 
running model achieves a better fit at low $\ell$. In Figure 
\ref{fig:axion_vs_run} we plot the primordial scalar power spectrum for axion 
monodromy models (solid lines) and constant running models (dashed lines), with 
$r$ fixed to 0.07 (dark lines) and 0.19 (red lines) in each case. For either 
$r$-value, we find that the constant running model achieves greater suppression 
of power at low $k$ (and hence, low $\ell$).  For the $r=0.19$ case we need a 
much greater suppression compared to $r=0.07$ to make up for the additional 
tensor power.  However, it is evident that the axion monodromy $r=0.19$ model
achieves much less suppression at low $k$ compared to the corresponding 
constant running model.  This occurs because the $r=0.19$ model has a 
relatively high exponent $p \approx 3$, as required by the e-folding constraint 
(eq.~\ref{efoldings_formula}).  As a result, the relatively large monomial term 
``softens'' the magnitude of the running at low $k$. The higher the $p$ (and 
hence, $r$), the more the running is mitigated at low $k$. This is the 
principle reason why axion monodromy provides a worse fit compared to constant 
running when $r$ is large.

Even for relatively small amplitudes, as $r$ is increased to ever higher 
values, $p$ must also be high (eq.~\ref{efoldings_formula}).  However, 
sufficiently high $p$ solutions lead to a very large or infinite number of 
e-foldings due to the appearance of a local minimum in the potential from the 
oscillation (unless the amplitude $b$ is very close to zero), as discussed in 
Section \ref{sec:priors}.  This effect excludes nearly all regions of parameter 
space where $p > 4$, and a large fraction of parameter space where $p > 3$.  
Thus, the direct effect of the oscillations on the total number of e-foldings 
$N$ restricts the allowed parameter space for large $r$ even further.

\begin{table*}
\centering
\begin{tabular}{|l|c|c|c|c|}
\hline
& Best-fit model & Best-fit constant $\alpha$ & Best-fit ($r=0.19$) & Best-fit constant $\alpha$ ($r=0.19$)\\
\hline
Lowlike (WMAP Pol.) & -0.5 & 0.1 & -1.1 & -1.0\\
Lensing & 0.8 & 0.9 & 1.0 & 1.2\\
Commander ($2 \leq \ell < 50$) & -1.8 & -2.5 & -0.6 & -1.7\\
CAMspec ($50 \lesssim \ell \lesssim 2500$) & 0.2 & 0.1 & 1.5 & 2.2\\
ACT/SPT ($600 \lesssim \ell \lesssim 3500$) & -0.5 & -0.2 & 0.2 & -0.1\\
\hline
Total & -1.7 & -1.6 & 1.0 & 0.6\\
\hline
\end{tabular}
\caption{$\Delta \chi^2$ compared to best-fit base Planck model (with $\alpha=0$)}
\label{tab:delta_chisq}
\end{table*}

\subsection{Comparison to constant running model at high multipoles}\label{sec:highl}

As Table \ref{tab:delta_chisq} shows, axion monodromy actually improves the fit 
for the ACT/SPT likelihood (high $\ell$) compared to the constant running 
model. We can understand why this is the case from the analytic expression for 
$n_s$ at the pivot scale $k=k_*$ (derived in appendix 
\ref{sec:ns_alpha_formulas}).  The spectral index $n_s$ has two contributions,

\begin{equation}
n_s \approx n_{s,0} + \Delta n_s,
\label{ns_total}
\end{equation}
where $n_{s,0}$ is the contribution from the monomial term in the potential 
(i.e.~the unmodulated spectral index), while $\Delta n_s$ is the contribution 
from the sinusoidal term.  By comparing to equation \ref{ns_formula} and 
substituting the approximate e-folding constraint (equation 
\ref{efoldings_formula_approx}), we find that

\begin{eqnarray}
n_{s,0} & = & 1 - \frac{r}{8}\left(1 + \frac{2}{p}\right) \label{ns0_exact} \\
& \approx & 1 - \frac{r}{8} - \frac{1}{N_0} \label{ns0_approx}
\end{eqnarray}
while for the oscillating term we have
\begin{equation}
\Delta n_s \approx -\frac{b}{f}\sqrt{\frac{r}{8}}\sin\delta + \frac{r}{4}\left(1-\frac{1}{p}\right)b\cos\delta.
\label{delta_ns}
\end{equation}

Generally, the first term in equation \ref{delta_ns} dominates over the second 
term provided that the phase shift $\delta$ is not very small or negative, 
since $b$ and $f$ are of a similar order. Given that the data prefers a 
positive $\delta$, we may therefore expect $\Delta n_s < 0$, and therefore 
$n_{s,0}$ should be at least slightly higher than $n_s$.

In Figure \ref{fig:post_derivedparams} we plot derived posteriors in $n_s$ and 
$n_{s,0}$ using equations \ref{ns_total}, \ref{ns0_exact}, and \ref{delta_ns}.  
The constraint in $n_s$ ($n_s \approx 0.96$) is quite similar to that obtained 
in the base Planck model (\citealt{planck2013}), which is expected since it is 
well constrained by the data. By contrast, the best-fit value for $n_{s,0}$ is 
$\approx 0.98$, which is strikingly high. This follows from equation 
\ref{ns0_approx}: since the number of e-foldings is restricted to the 
approximate range 50-60, the lower the $r$-value (and the corresponding 
$p$-value), the higher $n_{s,0}$ must be. Since the best-fit model has 
$r\approx 0.07$, this relatively low $r$ accounts for why $n_{s,0}$ is so high.

The question remains, why does the data at high $\ell$ prefer such a high 
$n_{s,0}$? According to Table \ref{tab:bestfit}, the best-fit axion monodromy 
model shows the greatest improvement in $\chi^2$ for the ACT/SPT likelihood. We 
have seen that at high multipoles, the data shows no strong preference for a 
running spectral index, with $n_s\approx 0.96$ preferred even at higher $\ell$.  
This is also largely true when the high-$\ell$ data from ACT/SPT is factored 
in, although it should be noted that ACT and SPT are in tension here; SPT 
prefers negative running while ACT does not (\citealt{valentino2010}).  Thus, 
while negative running improves the fit at low $\ell$, it worsens the fit at 
the high-$\ell$ end unless the running is relatively small.  In the best-fit 
axion monodromy model however, the running diminishes in magnitude as one 
proceeds to smaller scales (higher $\ell$), and hence the spectral index 
$\Delta n_s$ is slightly higher at high $\ell$.  This is why the relatively 
high value of $n_{s,0}$ is preferred, since it allows a smaller running at high 
$\ell$.

A similar conclusion regarding a high $n_{s,0}$ was reached by 
\cite{meerburg2014}, in which their best-fit solution has $n_{s,0} \approx 1$ 
for the case where they incorporate the BICEP2 constraint $r \sim 0.2$.  
However, in their work the full inflation model is not used; the spectral index 
$n_{s,0}$ is chosen completely independently of $r$, and no e-folding 
constraint is applied.  Equation \ref{ns0_approx} shows that for $r=0.2$, 
$n_{s,0}$ cannot be larger than 0.975, and this upper limit is only possible in 
the limit $N_0 \rightarrow \infty$.  From this it can be concluded that while 
the high-$\ell$ data prefers a high $n_{s,0}$, this is forbidden by the 
e-folding constraint unless $r$ is relatively small.  Indeed, the $r=0.19$ 
axion monodromy model in Table \ref{tab:bestfit} has $n_s \approx n_{s,0}$ and 
the fit is actually slightly worsened at the high-$\ell$ end.

\subsection{Can the number of e-foldings be amended to allow a higher 
tensor-to-scalar ratio?}\label{sec:amending_efoldings}

While the e-folding constraint disfavors high $r$-values, one might question 
whether our model might be amended so that the number of e-foldings satisfies 
the constraints even for high $r$.  In particular, our model assumes that the 
oscillation amplitude of the potential $a$ remains constant all the way to the 
end of inflation. For this reason, solutions with sufficiently large $r$ (and 
hence $p$) tend to dramatically inflate the number of e-foldings near the end 
of inflation, excluding these solutions.  However, if the oscillation amplitude 
were to diminish before the end of inflation, these high-$r$ solutions could in 
principle still satisfy the e-folding constraint.

Furthermore, from the point of view of the underlying microphysics in the axion 
monodromy model, there is no necessary reason to expect the modulation 
parameters to remain constant during inflation. On the contrary, these 
parameters are determined by the values of dynamical moduli fields (e.g.  
related to the size and shape of the compactified extra-dimensional manifold) 
which might well evolve with time as inflation progresses. Of course, if the 
oscillation amplitude were to increase with time, the resulting constraint on 
$r$ would be even stricter.  However it is reasonable to consider the case 
where the oscillation amplitude dies out well before the end of inflation. To 
consider this, we note that the approximate e-folding parameter $N_0$ 
(eq.~\ref{efoldings_formula}) gives the number of e-foldings without the direct 
contribution of the sinusoidal oscillation; further, by far the largest 
contribution of the oscillation to the number of e-foldings $N$ occurs at 
scales smaller than that probed by the CMB.  This is particularly the case for 
solutions with high $p$ (and hence, high $r$) values.  Therefore, we can use 
$N_0$ as a conservative estimate of the number of e-foldings in the case where 
the oscillations die out at scales smaller than that of the CMB.

\begin{figure}
	\includegraphics[height=0.7\hsize,width=0.9\hsize]{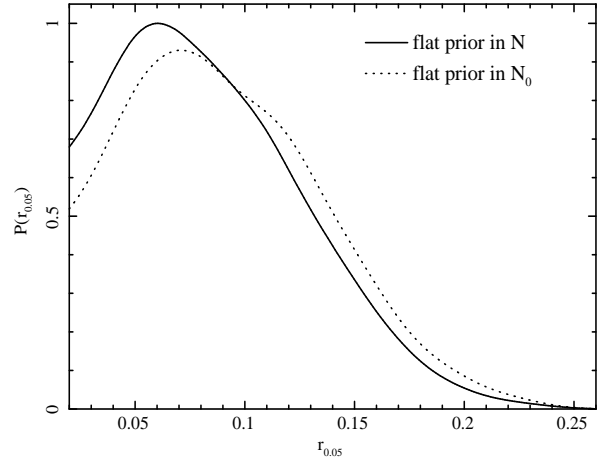}
\caption{Posterior in $r$ assuming two different e-folding priors. Solid line 
corresponds to a flat prior in $N$, which gives the number of e-foldings under 
the assumption the oscillation in the potential continued with undiminished 
amplitude to the end of inflation. Dotted line corresponds to a flat prior in 
$N_0$, which approximates the number of e-foldings in the scenario that the 
oscillation in the potential died out rapidly after the modes observed in the 
CMB exited the horizon. The former scenario gives smaller probability for large 
$r$-values, since it excludes regions of parameter space for which oscillations 
dominate near the end of inflation, resulting in a large or infinite number of 
e-foldings. However, $r=0.2$ is located outside the 99\% confidence region for 
the flat $N$ prior, and outside the 95\% confidence region for the flat $N_0$ 
prior and thus is disfavored in either scenario.}
\label{fig:post_r_N0}
\end{figure}

With this in mind, we apply a flat prior in the number of e-foldings $N_0 \in 
(50,60)$ to approximate the case where the amplitude dies out at smaller 
scales.  The resulting constraint on $r$ is shown in Figure 
\ref{fig:post_r_N0}, plotted next to the fiducial result which uses a flat 
prior in $N$.  As can be seen, there is greater probability for large $r$, but 
even in this case $r > 0.2$ is disfavored at the 95\% CL, for the same reason 
outlined in the previous section: the large amplitude (and low $p$) required to 
achieve enough suppression of power at low $\ell$ is forbidden by the e-folding 
constraint (eq.~\ref{efoldings_formula}).

In spite of the above difficulties, certainly there are ways to decrease the 
number of e-foldings to allow for higher $r$.  For example, if the exponent $p$ 
switches to a lower value at small scales, inflation would end more quickly and 
the number of e-foldings would be decreased. From the standpoint of 
single-field inflation this seems highly unnatural, although such a transition 
might naturally occur if multiple fields contribute to the effective inflaton 
potential (for an example of this scenario see \citealt{kobayashi2014}). From 
the Occam's razor point of view however, it seems more likely that the 
tensor-to-scalar ratio is simply smaller than the BICEP2 result suggests, 
particularly given the concerns about contamination by dust polarization.

\section{Can axion monodromy alleviate the small-scale problems of 
$\Lambda$CDM?}\label{sec:smallscale_probs}

\subsection{Implications for dwarf galaxy formation}\label{sec:dwarfs}

We have shown that oscillation in the power spectrum can allow for a 
significant tensor-to-scalar-ratio, alleviating the tension with the standard 
$\Lambda$CDM + power-law spectrum model at very large scales, while still 
satisfying the e-folding requirement.  However, there are also apparent 
departures from $\Lambda$CDM for small-scale structure which can be mitigated 
by the same mechanism. First, there is the ``missing satellites'' problem, 
which refers to the fact that cosmological dissipationless N-body simulations 
predict a much larger number of dwarf satellite galaxies around the Milky Way 
than are actually observed (\citealt{klypin1999}).

The second problem is that in the centers of baryon-poor galaxies the measured 
dark matter density is systematically lower than predicted by dark matter-only 
$\Lambda$CDM simulations. In many cases, such as in low surface brightness 
galaxies and field dwarfs with rotation (\citealt{kuzio2006}; 
\citealt{simon2005}; \citealt{adams2014}; \citealt{gentile2004}; 
\citealt{salucci2012}; \citealt{oh2011}), this lower density is due to the 
presence of a constant dark matter density core (the so-called ``core-cusp'' 
problem). The problem seems to extend to the lowest masses, and recent work 
with dwarfs in the Local Group (\citealt{boylan2012}; 
\citealt{tollerud2014}; \citealt{kirby2014}) and further away 
(\citealt{ferrero2012}) show that they are also systematically under-dense 
compared to simple expectations from $\Lambda$CDM, the so-called ``too big to 
fail'' problem (\citealt{boylan2011}).

It is possible that the simple $\Lambda$CDM expectations are incorrect and 
feedback from supernovae (\citealt{governato2012}), reionization 
(\citealt{bullock2000}), and other effects of star formation 
(\citealt{brooks2012}) all change these expectations dramatically.  There is no 
consensus regarding either of these two problems.  Reionization will prevent 
small satellites from being bright enough to be observed but it is unclear 
whether this by itself explains the luminosity distribution of the known 
satellites.  There are $\Lambda$CDM-based solutions for the second problem but 
none that alleviate the problem for all the different types of galaxies.

\begin{figure}
	\includegraphics[height=0.7\hsize,width=0.92\hsize]{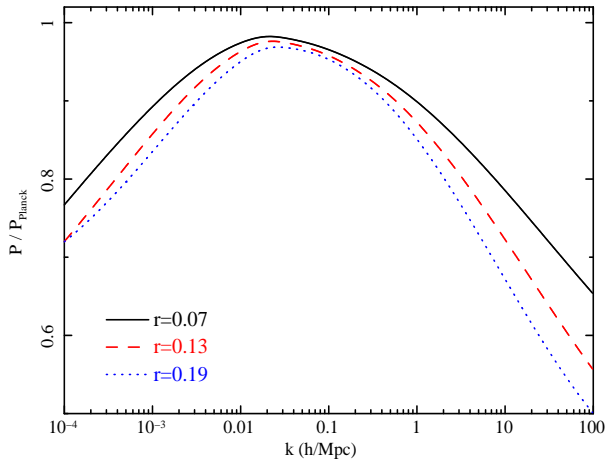}
\caption{The ratio of the matter power spectrum of axion monodromy compared to 
that of the base Planck model. Curves shown correspond to the best-fit model 
with the tensor-to-scalar ratio $r$ fixed to 0.07 (dark line), 0.13 (red dashed 
line), and 0.19 (blue dotted line). Note there is a reduction in power of 
$\sim$ 20-35\% at $k=10$ Mpc$^{-1}$, roughly the scale relevant to dwarf galaxy 
formation.}
\label{fig:relative_power}
\end{figure}

Here we ask whether the reduced power on small scales alleviates some of these 
issues. If the matter power spectrum is suppressed at the scales $k \sim 10$ 
Mpc$^{-1}$, then this would change the formation of dwarf galaxies. Halos would 
collapse at lower redshift, resulting in lower central dark matter densities.  
This reduced power may also change the way baryonic feedback operates in 
low-luminosity systems. This reduction in power has been considered in light of 
the BICEP2 result: if one allows a constant negative running $\alpha \approx 
-0.02$ in the power spectrum to allow for the high tensor-to-scalar ratio 
$r=0.2$ reported by BICEP2, \cite{garrison2014} found using N-body simulations 
that the too-big-to-fail problem is significantly alleviated, although not 
eliminated entirely (it should be noted that baryonic affects are not included 
in this work).  The reduction in power at dwarf scales in this ``BICEP2 
cosmology'' (\citealt{abazajian2014}) is approximately 40\% compared to vanilla 
$\Lambda$CDM.  However, as mentioned before, a constant running violates the 
e-folding constraint.  Since the model considered in this work achieves 
significant running with the requisite amount of e-foldings, we consider here 
the suppression of power at small scales from our model.  Again, we do not 
assume the BICEP2 result, but rather consider three different $r$-values 0.07, 
0.13, and 0.19.

In Figure \ref{fig:relative_power} we plot the relative power obtained by 
dividing the matter power spectrum in our model by the best-fit Planck power 
spectrum with zero running. In this plot we use the fitting functions of 
\cite{eisenstein1998} for the transfer function. For each $r$-value considered, 
we find the best-fit solution using the same method outlined in Section 
\ref{sec:results}, except with $r$ fixed to the given value. We plot the 
best-fit cases where $N$ lies in the fiducial range $50-60$. Note that for each 
case, the difference in power compared to the base Planck model comes not only 
from the primordial power spectrum parameters, but also because of differences 
in the transfer function which is sensitive to cosmological parameters, 
particularly $\Omega_m$ and $H_0$.  For $r=0.19$ we found that most of our 
solutions had $p>3$ and thus in many cases, a local minimum formed in the 
potential rendering the number of e-foldings infinite. We did find one solution 
in the desired range, which is shown in the figure.

At dwarf scales ($k\sim 10$ Mpc$^{-1}$), we find an 22\% reduction in power for 
$r=0.07$, 28\% reduction for $r=0.13$, and 33\% reduction for $r=0.19$. This is 
not as drastic a reduction as in the BICEP2 cosmology ($\approx$ 40\%), and 
recall in addition that $r=0.19$ is disfavored by the Planck data (Figure 
\ref{fig:fpriors}).  Nevertheless, a reduction in power of $\sim$ 20-30\% is 
entirely consistent with the data and e-folding constraint, and may be expected 
to alleviate the too-big-to-fail problem, particularly when considered in 
combination with baryonic feedback effects. One can also see from Figure 
\ref{fig:relative_power} that even greater suppression occurs at smaller 
scales, and this has a bearing on the missing satellites problem. In addition, 
the corresponding suppression of substructure would substantially reduce the 
expected dark matter annihilation signal (\citealt{garrison2014}).

\begin{figure}
	\includegraphics[height=0.85\hsize,width=0.92\hsize]{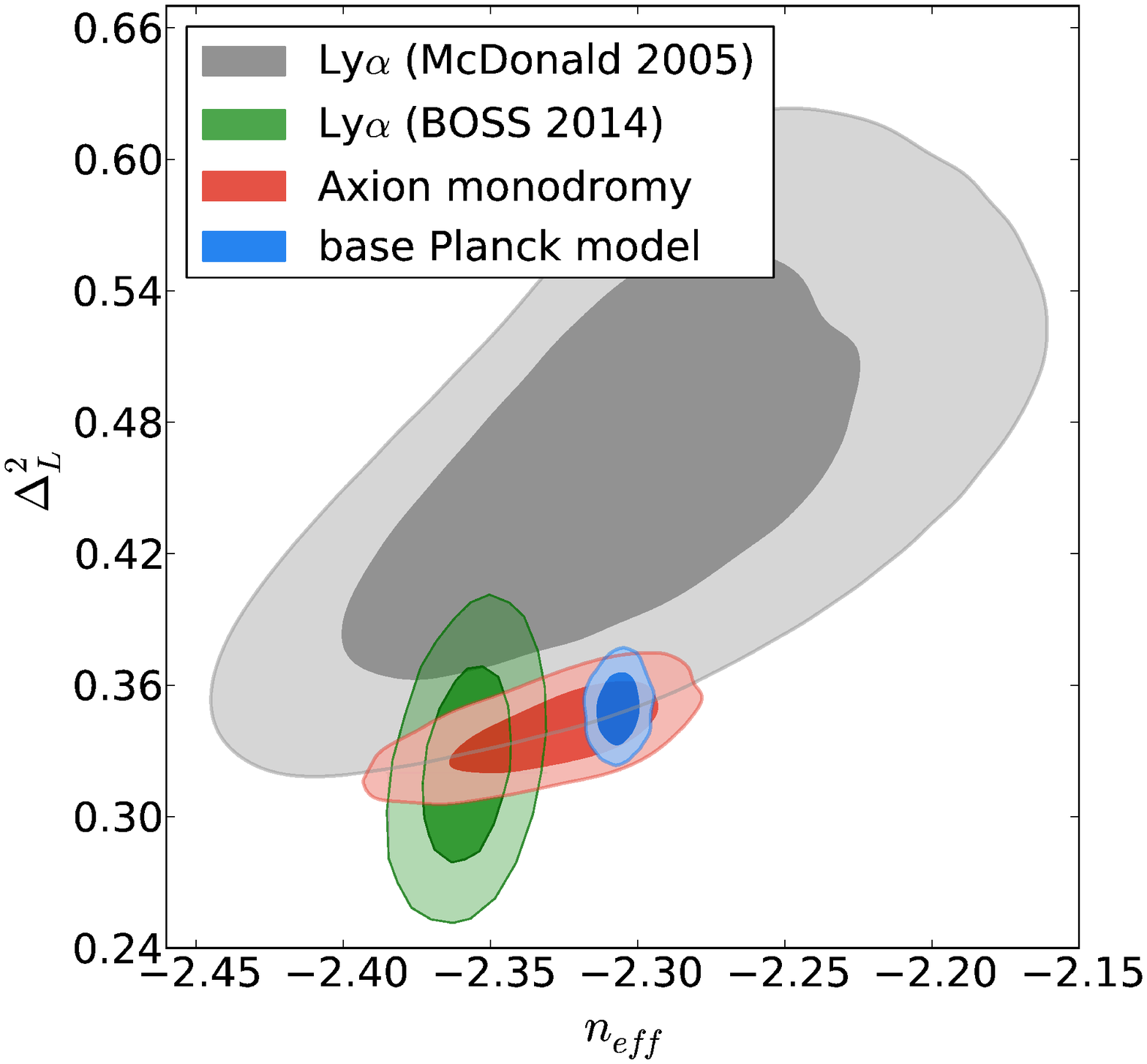}
\caption{Constraints on the amplitude $\Delta_L^2$ and slope $n_{eff}$ of the 
linear matter power spectrum at $z=3$ and $k=0.009$ (km/s)$^{-1}$ where $k$ is 
given in velocity space. Grey contours are from Lyman-$\alpha$ forest data 
analyzed in \cite{mcdonald2005}; green contours are from Lyman-$\alpha$ forest 
data from the BOSS survey (\citealt{boss2014a}); red and blue contours are from 
the marginalized posterior probability for the axion monodromy model and the 
base Planck model (with zero running), respectively. Note that the more recent 
BOSS data are in tension with the base Planck model at greater than 2-sigma, 
whereas they are consistent with axion monodromy to within 1-sigma.}
\label{fig:lyalpha}
\end{figure}

\subsection{Comparison to power spectrum constraints from Lyman-$\alpha$ forest 
data}\label{sec:lyman_alpha}

Aside from dwarf galaxies, another approach to detect oscillations in the power 
spectrum is to observe at high redshift where the matter power spectrum is 
close to linear at small scales.  The Lyman-$\alpha$ forest observed in quasar 
spectra can reveal structure down to $\sim$ 100 kpc scales in the approximate 
redshift range $z=2$ to $z=4$, which is in the quasi-linear regime.  Past 
studies have found constraints on the running of the spectral index, though not 
with the accuracy required for a positive detection of running of order $\alpha 
\approx -0.01$ (\citealt{seljak2006}; \citealt{viel2004}). However, more recent 
quasar observations by the Baryon Oscillation Spectroscopic Survey (BOSS) 
collaboration (\citealt{dawson2013}; \citealt{boss2014a}) have dramatically 
enlarged the available dataset of quasars with measured Ly$\alpha$ forest power 
spectra.  In Figure \ref{fig:lyalpha} we plot inferred probability contours in 
the amplitude $\Delta_L^2$ and slope $n_{eff}$ of the linear matter power 
spectrum at $z=3$ and $k=0.009$ km/s (corresponding to roughly 1 Mpc scales).  
The grey contours are from \cite{mcdonald2005}, while the green contours are 
from the more recent data from \cite{boss2014b}.  The red and blue contours are 
from derived posterior probability distributions for axion monodromy and the 
base Planck model respectively.  From this figure it is evident that while the 
older data cannot distinguish between the two models, the new BOSS data favors 
axion monodromy: the base Planck model prediction for $n_{eff}$ is more than 
2-sigma apart from that of \cite{boss2014b}, while the axion monodromy 
prediction is consistent to within 1-sigma. More generally, this comparison 
indicates that a negative running in the power spectrum may now be favored by 
recent Ly$\alpha$ forest data, at the level required to accommodate a 
substantial tensor-to-scalar ratio.

There are a number of other approaches to probing the small-scale power 
spectrum besides Lyman-$\alpha$ forest data. Down the road, 21-cm observations 
are a promising approach to constraining the power spectrum at even smaller 
scales (\citealt{shimabukuro2014}) since, unlike the Lyman-$\alpha$ forest, 
much higher redshifts can be probed where the data is not as limited at large 
$k$ by thermal broadening.  In the meantime, resolving the discrepancy between 
SPT and ACT on whether negative running is preferred in the CMB at high $\ell$ 
will be an important step.  Within the coming decade, advanced redshift surveys 
such as WFIRST (\citealt{spergel2013}) and LSST may offer the best window onto 
the power spectrum at scales smaller than CMB (\citealt{meerburg2014}), with 
the caveat that galaxy bias must be properly taken into account. In any event, 
a non-trivial primordial power spectrum remains an intriguing possibility for 
alleviating small-scale problems. This will be true particularly if alternate 
forms of dark matter (WDM, self-interacting) are ruled out as solutions to the 
too-big-to-fail and/or missing satellites problems.

\section{Conclusions}\label{sec:conclusions}

In this paper we have investigated whether inflation models can allow for a 
large gravitational wave background while remaining consistent with the 
CMB power spectrum at horizon scales (low $\ell$).  We have shown that axion 
monodromy inflation accomplishes this naturally through Planck-scale 
corrections, producing a gentle oscillation in the inflaton potential 
(eq.~\ref{potential_original}).  This generates a running spectral index in the power 
spectrum while still achieving enough e-foldings to solve the horizon problem, 
in stark contrast to the usual constant running model.  We have fit our model to a 
combination of Planck, ACT, SPT, and WMAP low-$\ell$ polarization CMB data 
together with a prior on the number of e-foldings. The best-fit model 
parameters are given in Table \ref{tab:bestfit}, while inferred probability 
distributions in the inflation model parameters are shown in Figure 
\ref{fig:triangle_plot}.

We find a best-fit tensor-to-scalar ratio $r = 0.07^{+0.05}_{-0.04}$ and thus 
the predicted imprint of gravitational waves on the CMB is within reach of 
B-mode polarization experiments. It is possible that these
primordial B-modes have already been observed by the BICEP2 experiment; however 
if one assumes neglible foreground contamination, the BICEP2 result
$r\approx0.2$ is disfavored at the 99\% confidence level. This is primarily a 
consequence of the e-folding constraint and the requirement to fit the spectral 
index at high $\ell$ in addition to the low-$\ell$ power spectrum.  Since a 
running spectral index is the most straightforward way to reconcile the BICEP2 
result with Planck, it is significant that attempting to implement running in 
the underlying inflation theory disfavors such a large $r$, as we have shown 
here.  While it is possible that a dramatic change in the inflaton potential at 
small scales (high $k$) can alter the e-folding constraint to allow for a 
higher $r$, we find it more likely that the tensor-to-scalar ratio is simply 
smaller than the best-fit BICEP2 result suggests, particularly in light of the 
uncertainties about foreground contamination by dust polarization in the BICEP2 
field.

In addition to the large gravitational wave amplitude, the (best-fit) axion 
monodromy model makes two corresponding predictions: first, despite the 
additional tensor power on horizon scales, the overall power at low 
multipoles is reduced as a consequence of the running spectral index, providing 
a better fit to the CMB power spectrum at large scales.  The second prediction 
is that the matter power spectrum is suppressed at the scale of dwarf galaxies, 
and thus axion monodromy can alleviate some of the small-scale problems of 
$\Lambda$CDM---in particular, the too-big-to-fail and missing satellites 
problems.  We find that our best-fit models reduce the power at scales relevant 
to dwarf galaxy formation ($k \sim 10$ Mpc$^{-1}$) by as much as $\sim$35\% 
depending on the assumed $r$-value, with the greatest reduction achieved at 
large $r$.  However, 20-30\% suppression is more likely, which will alleviate 
the too-big-to-fail problem and may solve it entirely when combined with 
baryonic feedback effects, as discussed in \cite{garrison2014}. Additionally, 
we find that axion monodromy is preferred by recent Ly$\alpha$ forest data over 
the base Planck model without running (Figure \ref{fig:lyalpha}).

If axion monodromy (or a similar oscillating large-field model) accounts for 
the reduced power at large and small scales, then the tensor-to-scalar ratio is 
likely to lie in the range $r\approx 0.03-0.12$ (68\% CL) and hence the imprint 
of gravitational waves on the CMB will be observable by B-mode experiments in 
the future (\citealt{abazajian2013}). With this comes the tantalizing prospect 
of constraining physics at the Planck scale through sky surveys, as we have 
demonstrated here. Thus, future CMB experiments, in combination with probes of 
the power spectrum at small scales, may settle the issue of whether 
Planck-scale physics manifest in inflation can reconcile the standard 
$\Lambda$CDM cosmology with data at all observable scales of the Universe.

\section*{Acknowledgements}
QM would like to thank James Bullock, Shahab Joudaki, Jose Ceja and Shea 
Garrison-Kimmel for their encouragement and feedback at the 
beginning stages of this project.

We gratefully acknowledge a grant of computer time from XSEDE allocation 
TG-AST130007. MK was supported in part by NSF grant PHY-1214648.

This research was also supported, in part, by a grant of computer time from the 
City University of New York High Performance Computing Center under NSF Grants 
CNS-0855217, CNS-0958379 and ACI-1126113.

\bibliography{monodromy}

\setcounter{section}{0}
\section{Appendix: Is the slow roll approximation valid in our 
model?}\label{sec:slowroll}

When there are oscillating features in the power spectrum, the slow roll 
approximation can be violated, particularly for short-period oscillations.  
Thus, it is important to check that the slow roll approximation we have used 
here is valid in the parameter range we are interested in. For the slow roll 
approximation to hold, both the first slow roll parameter $\epsilon = 
\frac{d}{dt}\left(\frac{1}{H}\right)$ and the second slow roll parameter $\eta 
= \frac{1}{H}\frac{\ddot\phi}{\dot\phi}$ must be small, such that $\epsilon \ll 
1, \eta \ll 1$. The former condition is obviously satisfied at CMB scales, as 
we are many e-foldings before the end of inflation and thus $\epsilon \ll 1$.  
The second condition is trickier, however, because rapid oscillation can cause 
$\eta$ to become large.  To estimate $\eta$, we must solve the classical 
equation of motion for $\phi(t)$:

\begin{equation}
\ddot\phi + 3H\phi + V_{,\phi} = 0
\end{equation}

Since $\epsilon$ is small, we have $H \approx \sqrt{\frac{V}{3}}$. Let us 
switch variables to $x=t\sqrt{\frac{V_*}{3}}$, where $x$ can be interpreted as 
the time in units of the Hubble time when the mode $k=k_*$ leaves the horizon.  
We then find

\begin{equation}
\frac{d^2\phi}{dx^2} + 3\sqrt{\frac{V}{V_*}}\frac{d\phi}{dx} + 3\frac{V_{,\phi}}{V_*} \approx 0.
\label{phi_evolution}
\end{equation}

This equation can be solved numerically, using the expression for the (scaled) 
potential and its derivative from equation \ref{potential}. Here, since we are 
interested in an analytic expression for $\eta$, we will use a method similar 
to that of \cite{flauger2010} and try a solution of the form $\phi = \phi_0 + 
a\phi_1$, where $\phi_0$ is the unmodulated scalar field, $\phi_1$ contains the 
oscillation and the amplitude $a$ is assumed to be small (as it must be). For 
$\phi_0$, the slow roll approximation holds very well, as usual, so we can 
ignore $\ddot\phi_0$.  Substituting this expression in to equation
\ref{phi_evolution} and separating out the zeroth order and first order terms, 
we find for the zeroth order equation

\begin{equation}
\frac{d\phi_0}{dx} = \frac{p}{\phi_{min}}\left(1-\frac{\phi_0}{\phi_{min}}\right)^{\frac{p}{2}-1}.
\label{dphidx}
\end{equation}
We will not need to solve this equation exactly for the present discussion, but 
we note that since we are 50-60 e-foldings from the end of inflation, we have 
$\phi_0 \ll \phi_{min}$ and hence, $\frac{d\phi_0}{dx} \approx 
\frac{p}{\phi_{min}}$.

Proceeding to the first-order equation, we switch to using $\phi_0$ as the 
independent variable, yielding the equation
\begin{equation}
\frac{1}{3}\phi_1'' + \frac{1}{p}\phi_{min}\phi_1' - \frac{1}{2}\phi_1 = -\frac{\phi_{min}^2}{p^2f}\cos\left(\frac{\phi_0}{f} + \delta\right).
\end{equation}
where the prime denotes the derivative $\partial/\partial\phi_0$. To solve this 
equation we try a linear combination of sines and cosines, and arrive at the 
solution

\begin{equation}
\phi_1(x) \approx \frac{3f\left(\frac{\phi_{min}}{nf}\right)^2}{\sqrt{\left(1+\frac{3}{2}f^2\right)^2 + \left(\frac{3\phi_{min}f}{p}\right)^2}}\cos\left(\frac{\phi_0(x)}{f} + \psi\right).
\label{phi1}
\end{equation}
where $\psi$ is an unimportant phase factor. We can now estimate the amplitude 
of the slow roll parameter $\eta = \frac{1}{H}\frac{\ddot\phi}{\dot\phi} 
\approx a\frac{d^2\phi_1}{dx^2} / \frac{d\phi_0}{dx}$. Plugging in 
eqs.~\ref{dphidx} and \ref{phi1}, we find

\begin{equation}
\eta_{max} \approx \frac{3a\frac{\phi_{min}}{nf}}{\sqrt{\left(1+\frac{3}{2}f\right)^2 + \left(\frac{3\phi_{min}f}{p}\right)^2}}.
\label{eta_max}
\end{equation}

This is well approximated as
\begin{equation}
\eta_{max} \approx \frac{3b}{2+b\cos\delta}\left(1+18N_0\frac{f^2}{p}\right)^{-1/2},
\end{equation}
where we have made use of equations \ref{bdef}, \ref{phimin}, and 
\ref{phimin_efoldings}. Note that if the amplitude $b$ is kept fixed, $\eta$ 
can be made large (and hence, slow roll violated) if $f$ is sufficiently small.  

\begin{figure}
	\centering
	\includegraphics[height=0.4\hsize,width=0.53\hsize]{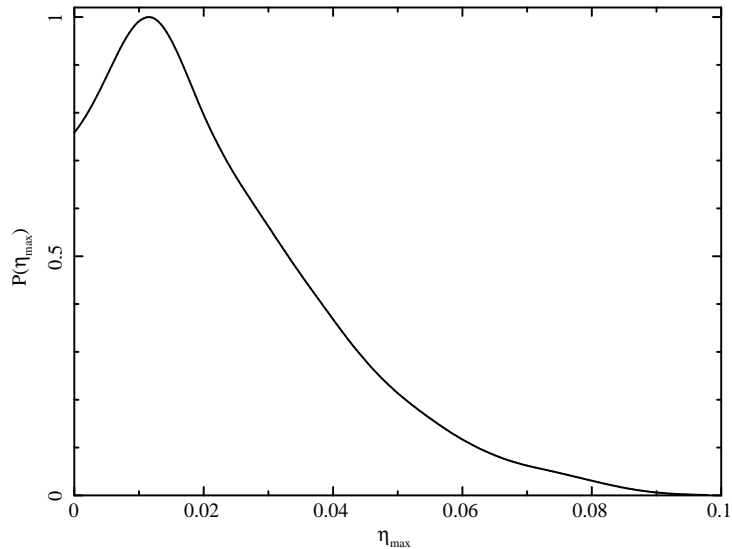}
\label{fig:etamax}
\caption{Posterior in the amplitude of the slow roll parameter $\eta$. Since 
$\eta_{max} \ll 1$ for all allowed regions of parameter space, the slow roll 
expansion holds to good approximation, justifying the method used to derive the 
power spectra in Section \ref{sec:power_spectrum}.}
\end{figure}

For the best-fit solution in Table \ref{tab:bestfit}, we have $\eta_{max} 
\approx 0.04$ and thus the slow roll description holds. On the other hand, at 
the extreme ends of the parameter space we have chosen, $\eta$ can become 
large; for example, setting $b=1$, $f=0.1$ can give $\eta_{max}$ comparable to 
1.  Fortunately, such regions where slow roll breaks down are excluded by the 
data.  To show this robustly, in Figure \ref{fig:etamax} we plot a derived 
posterior in $\eta_{max}$ using equation \ref{eta_max}, with the same priors 
and data as discussed in Section \ref{sec:priors}. For all the points sampled, 
we have $\eta_{max} \lesssim 0.1$, and hence the slow roll approximation holds 
good for all allowed regions of parameter space.  This justifies our use of the 
slow roll approximation in deriving the power spectra in Section 
\ref{sec:power_spectrum}.  \\

\section{Appendix: Expressions for the spectral index and running in the slow-roll approximation}\label{sec:ns_alpha_formulas}

To relate the parameters to observable features, it is useful to have 
expressions for the spectral index and running at the pivot scale. The most 
useful expressions we derive will be equations \ref{ns_formula} and 
\ref{alpha_formula}, which are given in terms of the fit parameters in the usual 
case where $a \ll 1$, and the running is dominated by the sinusoidal term.

\subsection{Spectral index}

To derive expressions for the spectral index, we can use the relation in terms 
of the slow roll parameters, $n_s = 1 - 6\epsilon_V + 2\eta_V$. Using the fact 
that $\epsilon_V = r/16$ and $\eta_V=V_{,\phi\phi}/V$, we find using equation 
\ref{potential},

\begin{equation}
n_s = 1 - \frac{3r}{8} + 2\left[\frac{p(p-1)}{\phi_{min}^2}(1-a\sin\delta) - \frac{a}{f^2}\sin\delta\right]
\end{equation}
where it is understood that we are evaluating the spectral index at the pivot 
scale $k=k_*$. Substituting equation \ref{r_constraint} and keeping terms to first 
order in $a$, we find (after substituting equation \ref{bdef}),

\begin{equation}
n_s ~ \approx ~ 1 - \frac{r}{8}\left(1 + \frac{2}{p}\right) - \frac{b}{f}\sqrt{\frac{r}{8}}\sin\delta + \frac{r}{4}\left(1-\frac{1}{p}\right)b\cos\delta.
\label{ns_formula}
\end{equation}
Note that the oscillation is dominated by the sine term (unless $\delta$ is 
very small such that $\tan\delta \lesssim f\sqrt{\frac{r}{8}}$), and thus the 
amplitude of the spectral index oscillation is $\Delta n_s \approx 
\frac{2a}{f^2} \approx \frac{b}{f}\sqrt{\frac{r}{8}}$. Our constraint in $n_s$ 
is plotted in Figure \ref{post_ns} and is consistent with the result $n_s 
\approx 0.96$ found by applying the base Planck model.

\subsection{Running of the spectral index}

As discussed in Section \ref{sec:results}, the best fit parameters are 
characterized by a substantial running of the spectral index which is dominated 
by the sinusoidal term. To first order, the running can be calculated from 
$\alpha_1 \approx - 2\zeta_V$ where $\zeta_V = 
\frac{V_{,\phi}V_{,\phi\phi\phi}}{V^2} = 
\sqrt{2\epsilon_V}\frac{V_{,\phi\phi\phi}}{V}$ is the third potential slow roll 
parameter.  From this we find (at the pivot scale $k_*$),

\begin{equation}
\alpha_{1,*} \approx -\frac{rb}{8f^2}\cos\delta.
\label{alpha_approx}
\end{equation}

From this formula it is evident that the running oscillates in $\phi \propto 
\ln k$ with the approximate amplitude $\alpha_{max} \approx \frac{rb}{8f^2}$.

Now we derive the more exact expression for the running using the formula 
$\alpha = -2\zeta_V + 16\epsilon_V\eta_V - 24\epsilon_V^2$.  Expanding this to 
first order in $a$ and up to second order in $r$, we obtain

\begin{equation}
\alpha_* \approx -\frac{rb}{8f^2}\cos\delta - \frac{4b}{f}\left(\frac{r}{8}\right)^{3/2}\sin\delta + \frac{r^2}{8}\left(1-\frac{1}{p}-\frac{3}{4}\right).
\label{alpha_formula}
\end{equation}

The last term above is due to the monomial term in the potential, and is 
clearly of order $10^{-3}$ if $r \sim 0.1$; the middle term is typically of a 
similar order.  For the best fits, we have $\alpha \sim -10^{-2}$, in which 
case the above formula is accurate for up to two significant figures. Our 
constraint in the running $\alpha_*$ is plotted in Figure \ref{post_alpha}.

\begin{figure*}
	\centering
	\subfigure[]
	{
		\includegraphics[height=0.28\hsize,width=0.31\hsize]{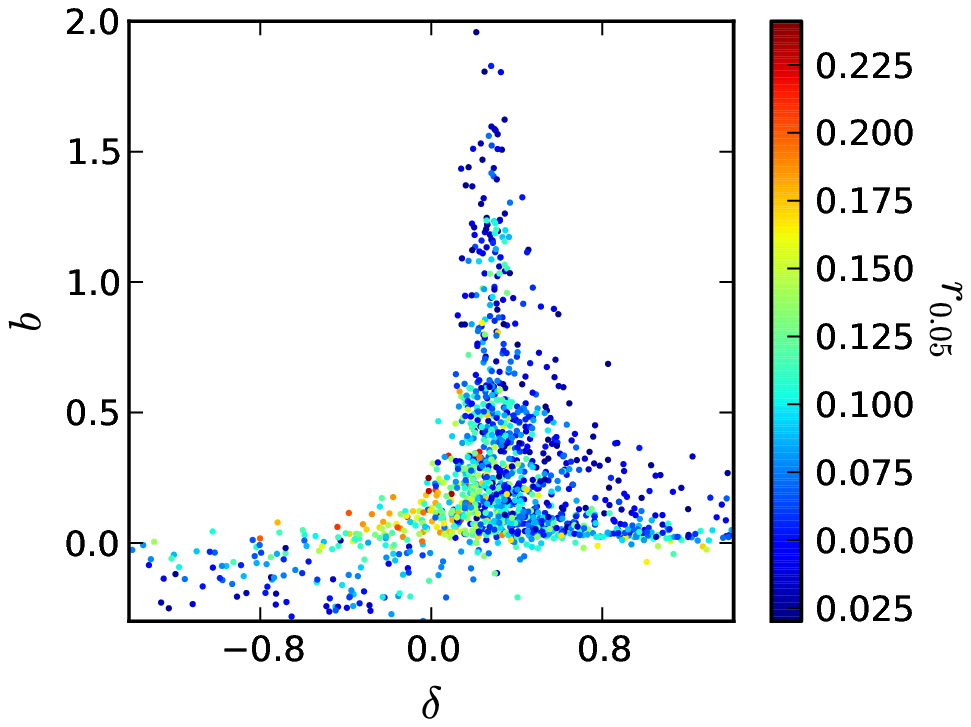}
		\label{r_vs_bd}
	}
	\subfigure[]
	{
		\includegraphics[height=0.28\hsize,width=0.31\hsize]{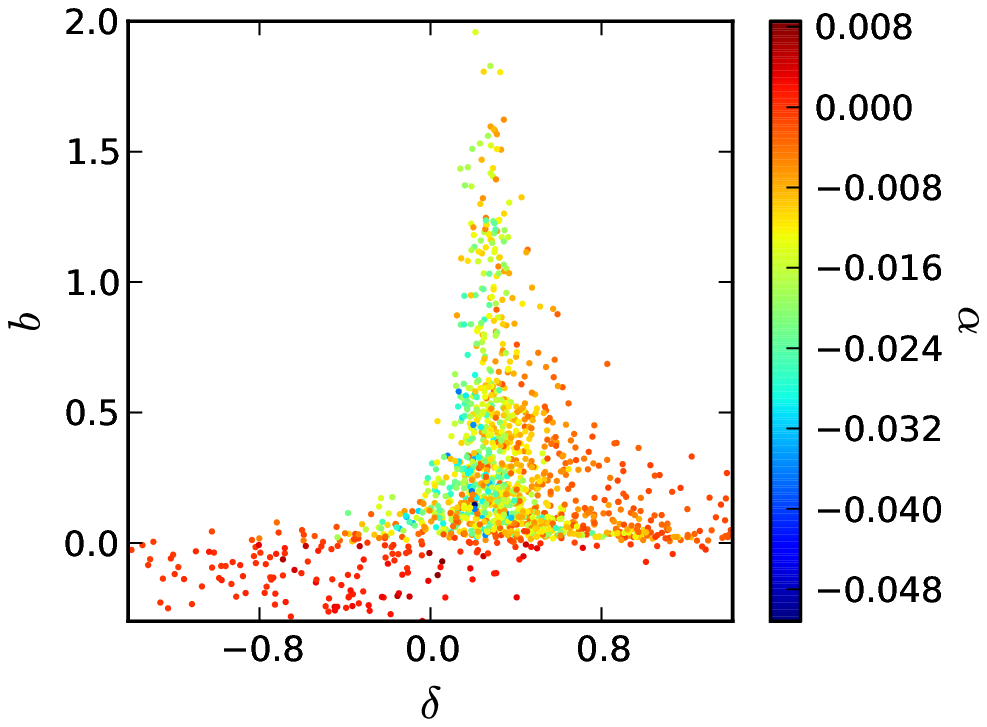}
		\label{alpha_vs_bd}
	}
	\subfigure[]
	{
		\includegraphics[height=0.28\hsize,width=0.31\hsize]{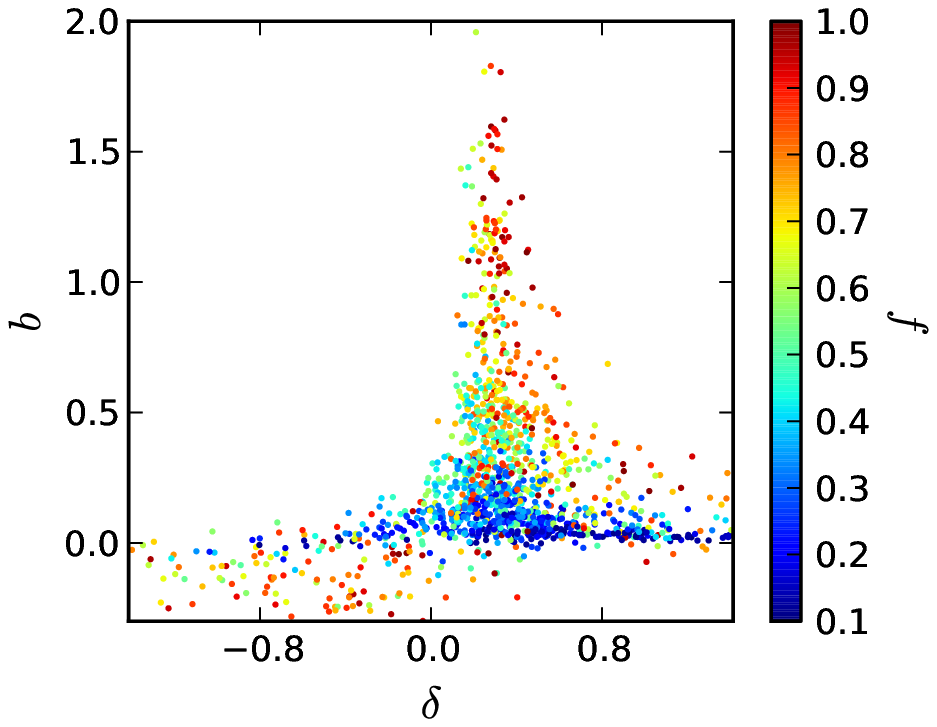}
		\label{f_vs_bd}
	}
	\caption{Joint 3-dimensional posteriors in the amplitude $b$ and phase shift 
$\delta$, color-coded by the tensor-to-scalar ratio $r$, running $\alpha_*$, 
and axion decay constant $f$ respectively.  Note that outside the best-fit 
region, there are tails extending to large $b$, or small $b$ and extreme values 
of $\delta$.  In these tails, $r$ (and correspondingly, the running $\alpha_*$)
prefers to be small. Likewise, $f$ tends to take extreme values in the tails 
while preferring intermediate values in the best-fit region.}
	\vspace{10pt}
\label{fig:3dposts}
\end{figure*}

\section{Appendix: Structure of the posterior distribution}\label{sec:3dposts}

As discussed in Section \ref{sec:results}, the posterior distribution has a 
complicated multi-modal structure, with several non-Gaussian tails running away 
from the best-fit region. To understand this structure better, in Figure 
\ref{fig:3dposts} we plot 3-dimensional posteriors with the axes for $b$ and 
$\delta$ shown, color-coded by $r$, $\alpha_*$, and $f$ respectively. Note 
there are three separate ``wings'' in the distribution running away from the 
best-fit region, and in all of these wings the tensor-to-scalar ratio $r$ 
prefers to be small (Figure \ref{r_vs_bd}). This is because large $r$ must be 
accommodated by a large negative running, and in Figure \ref{alpha_vs_bd} one 
sees that a large negative running only occurs in the best-fit region, whereas 
it is quite small in the wings. Two of the wings have $b \approx 0$, which then 
naturally would imply a small running; however, the other wing has a large 
amplitude $b \gtrsim 1$. How can the running be small in this case? The answer 
can be seen in Figure \ref{f_vs_bd}: in most of this region $f$ (corresponding 
to the period) is quite large, and thus the running is again small. Note that 
in all the wings $f$ tends to take on either very small ($f \approx 0.1$) or 
very large values ($f \approx 1$), with intermediate values occurring mostly in 
the best-fit region. For this reason, if one stays within the best-fit region, 
the axion decay constant $f$ is somewhat better constrained than is suggested 
by the posterior in Figure \ref{fig:triangle_plot} although the errors are 
still large.

\end{document}